\documentclass[fleqn,usenatbib]{mnras}

\usepackage{newtxtext,newtxmath}

\usepackage[T1]{fontenc}

\DeclareRobustCommand{\VAN}[3]{#2}
\let\VANthebibliography\thebibliography
\def\thebibliography{\DeclareRobustCommand{\VAN}[3]{##3}\VANthebibliography}

\usepackage{graphicx}	
\usepackage{amsmath}	
\usepackage{textcomp}	
\usepackage[dvipsnames]{xcolor}
\usepackage{tikz}
\usetikzlibrary{arrows.meta,
                 shapes,
                chains,
                positioning,
                shapes.geometric}
\tikzstyle{arrow}=[draw, -latex]

\newcommand{\appropto}{\mathrel{\vcenter{
  \offinterlineskip\halign{\hfil$##$\cr
    \propto\cr\noalign{\kern2pt}\sim\cr\noalign{\kern-2pt}}}}}

\newcommand{\cs}{c_\mathrm{s}}

\newcommand{\Mp}{M_\mathrm{p}}

\newcommand{\Mth}{M_\mathrm{th}}
\newcommand{\hp}{h_\mathrm{p}}

\newcommand{\Rp}{R_\mathrm{p}}
\newcommand{\bfR}{{\bf R}}
\newcommand{\bfRp}{{\bf R}_\mathrm{p}}
\newcommand{\aid}{\textbf{\textit{a}}_\mathrm{id}}

\newcommand{\Omegap}{\Omega_\mathrm{p}}

\newcommand{\de}{\mathrm{d}}
\newcommand{\M}{\mathcal{M}}

\newcommand{\I}{\mathrm{i}}
\newcommand{\e}{\mathrm{e}}

\newcommand{\IM}{\mathrm{Im}}

\newcommand{\Pp}{P_\mathrm{p}} 

\newcommand\ba{\begin{eqnarray}}
\newcommand\ea{\end{eqnarray}}

\defcitealias{Goodman2001}{GR01}
\defcitealias{Rafikov2002}{Rafikov 2002}
\defcitealias{Cimerman2021}{Paper I}
\defcitealias{GT80}{GT80}

\title[Indirect forces]{Indirect forces in disc-planet interaction}

\author[R. R. Rafikov]{Roman R. Rafikov$^{1,2}$\thanks{E-mail: rrr@damtp.cam.ac.uk}, Nicolas P. Cimerman$^{1}$, Callum W. Fairbairn$^{2}$, \newauthor \& Alexander J. Dittmann$^{2}$\thanks{NASA Einstein Fellow}\\
$^{1}$Department of Applied Mathematics and Theoretical Physics, University of Cambridge, Wilberforce Road, Cambridge CB3 0WA, UK\\
$^{2}$Institute for Advanced Study, Einstein Drive, Princeton, NJ 08540, USA}

\date{Accepted XXX. Received YYY; in original form ZZZ}

\pubyear{2025}

\begin{document}

\def\etal{et al.\ \rm}
\def\etal{et al.\ \rm}
\def\Fdw{F_{\rm dw}}
\def\Fdis{F_{\rm dw,dis}}
\def\Fnu{F_\nu}
\def\WD{\rm WD}

\label{firstpage}
\pagerange{\pageref{firstpage}--\pageref{lastpage}}
\maketitle

\begin{abstract}
Gravitational coupling between a protoplanetary disc and an embedded planet is often studied in a frame attached to a central star. This frame is non-inertial because of the stellar reflex motion, leading to indirect forces arising in the star-planet-disc system. Here we examine the impact produced by these forces on several aspects of disc-planet coupling using analytical and numerical means. We explore how neglecting indirect forces changes (1) the spatial pattern of the surface density perturbation in the disc, (2) the calculation of the torque exerted on the disc by the planet, and (3) the torque on the planet exerted by the disc.  For low-mass planets, in the linear regime, the differences in the perturbation pattern are only in its $m=1$ azimuthal harmonic, with an amplitude increasing with the distance from the star. In this regime both the torque on the planet and the deposition torque density in the disc are only weakly affected by non-inclusion of indirect forces, corroborating some results of studies neglecting indirect forces altogether. For higher mass planets, a broader range of azimuthal harmonics of the perturbation are affected. Also, indirect forces have a stronger effect on the planetary torque and on planet migration in the Type II regime. We highlight the importance of including the planetary indirect force in the calculation of the torque on the disc (if disc evolution accounts for indirect force) to ensure conservation of angular momentum carried by the planet-driven density waves. The corresponding indirect torque has an oscillatory, radially-diverging character.
\end{abstract}





\section{Introduction}
\label{sec:intro}


Gravitational coupling between a planet orbiting a central star and a protoplanetary disc in which the planet may be embedded is an important astrophysical problem \citep{GT80}. Studying it involves considering a dominant mass, a star, that is orbited by a low-mass object, a planet, with both acting gravitationally on a disc. In this case it is natural to consider disc and planetary dynamics in the frame co-moving with the star since the inner parts of the disc (lying interior to the orbit of the planet) tend to move on Keplerian orbits centered on the star. 

It is also well known that attaching the reference frame to one of the objects in a system of many bodies gives rise to so-called {\it indirect forces} in the description of the motion of other bodies relative to the reference object. These forces account for the non-inertial motion of the reference object, i.e. the reflex motion of a star around the barycenter of the system in the case of disc-planet coupling. Indirect forces affect the dynamics of both the disc and the planet, potentially impacting our understanding of gap opening, planetary migration, and so on. 

There are multiple ways in which indirect forces affect disc-planet interaction. It is well known \citep{GT80} that this gravitational coupling is mediated by the excitation of a density wave that propagates through the disc carrying angular momentum and energy. In the absence of dissipation, the torque exerted by the planet on the disc is entirely carried away by the wave \citep{Papaloizou1984,Gold1989}, with the disc ffluid elements experiencing no net loss or gain of the angular momentum. But when dissipation, linear or nonlinear, is present, the wave damps and transfers its angular momentum to the disc, driving its evolution and leading to gap opening. In turn, the torque exerted by the perturbed disc on the planet results in the planetary orbit changing, i.e. orbital migration and eccentricity and inclination evolution. As we will see below, indirect forces enter at some level into the description of all of these phenomena --- the effect a planet has on the disc and the back-reaction of the disc on the planetary orbit.

Despite of this, many studies of disc-planet interaction explicitly neglect indirect forces when exploring the perturbation imposed by the planet on a disc or when studying the gravitational back-reaction of the disc on the planetary motion. This approach is typically based on the premise that the indirect forces should not strongly affect a particular aspect of the disc-planet problem and can thus be neglected. However a clear quantitative justification of this expectation is usually not provided, with rare exceptions \citep{RegalyVorobyov2017,Fairbairn2025a,Crida2025}.

The goal of this work is to provide a quantitative characterization of the impact of indirect forces on different aspects of the disc-planet interaction. Upon presenting our physical setup and numerical methodology in Section \ref{sec:setup}, we focus on the three key ingredients of the problem. First, we provide a quantitative measurement of the differences in the planet-driven disc perturbation pattern computed with and without the indirect forces (Section \ref{sec:pert}). Second, we study the impact of indirect forces on the key characteristics of the planet-driven density waves --- excitation and deposition torque densities and wave angular momentum flux --- and show that neglecting indirect forces can lead to paradoxical results (Section \ref{sec:IT-AM}). Third, we quantitatively examine the role of indirect forces in modifying the rate of planetary migration\footnote{This aspect of disc-planet coupling has been explored to some degree in the high planetary mass regime by \citet{RegalyVorobyov2017} and \citet{Crida2025}; we discuss these studies in Section \ref{sec:disc}.} (Section \ref{sec:mig}).


\section{Disc-planet setup and methodology}
\label{sec:setup}


We illustrate the effect of indirect forces in disc-planet coupling using a simple physical setup. We consider a system of a star of mass $M_\star$ and a planet of mass $\Mp$, in circular orbits around their common barycenter. Coplanar with the star-planet system is a thin (two-dimensional, 2D) gaseous disc. Considering the dynamics in the frame centered on the star, we will use $\bfRp$ as the position vector of the planet relative to the star and $\bfR$ as the position vector of a fluid element of the disc (also relative to the star). The mass distribution in the disc is characterized by the surface density $\Sigma(\bfR)$, which is in general non-axisymmetric, e.g. due to the planetary perturbation.


\subsection{Mathematical foundations}
\label{sec:math}


Equations governing the motion of $N>2$ gravitating objects (point masses) in a non-inertial reference frame attached to one of them can be found in many textbooks \citep[e.g.][]{Tremaine2023}. They can be easily generalized to a system which includes a continuous massive disc. Focusing on the motion of a disc fluid element at the position vector $\bfR$ in a frame attached to the central star, the governing equation including all indirect terms becomes \citep[e.g.][]{Lin2011}
\begin{align}
\ddot{\bf R}=&-GM_\star\frac{\bfR}{R^3}
-G\Mp \left(\frac{\bfR-\bfRp}{\vert\bfR-\bfRp\vert^3}+\frac{\bfRp}{R_\mathrm{p}^3}\right)
\nonumber\\
&- G\int\limits_{\rm disc}
\Sigma(\bfR^\prime)
\left(\frac{\bfR-\bfR^\prime}{\vert\bfR-\bfR^\prime\vert^3}+\frac{\bfR^\prime}{R^{\prime 3}}\right)\de^2 \bfR^\prime.
\label{eq:gasEoMRgen}
\end{align}
The different terms on the right-hand side describe (in order of appearance) stellar gravity, direct gravity and indirect force due to the planet, and direct self-gravitational and indirect effects from the rest of the disc (the integral is carried over the entire disc surface). Here (and in other equations) we omitted for simplicity all other forces except direct gravitational and indirect forces; however, our subsequent numerical calculations fully account for the pressure forces (see Section  \ref{sec:physical}). 

Similarly, the equation of motion for the planet (located at $\bfRp$) including the gravitational effect of the disc becomes 
\begin{align}
\ddot{\bf R}_\mathrm{p}  = & -G(M_\star+\Mp)\frac{\bfRp}{\Rp^3}
\nonumber\\
&- G\int\limits_{\rm disc}
\Sigma(\bfR^\prime)
\left(\frac{\bfRp-\bfR^\prime}{\vert\bfRp -\bfR^\prime\vert^3}+\frac{\bfR^\prime}{R^{\prime 3}}\right)\de^2 \bfR^\prime.
\label{eq:plEoMRgen}
\end{align}
The second line of this equation represents the gravity and indirect force due to the disc, as in the second line of (\ref{eq:gasEoMRgen}). 

The first term on the right hand side of (\ref{eq:plEoMRgen}) is usually interpreted as the acceleration caused by the `effective two-body force' between the star and the planet. However, as shown in \citet{Rafikov2025}, it is more physically appropriate to consider it as a direct gravitational acceleration due to the star (a part proportional to $M_\star$) and the indirect acceleration due to the planet (a part scaling with $\Mp$). Indeed, the full indirect acceleration, 
\begin{align}
\aid=-G\Mp\frac{\bfRp}{\Rp^3}
- G\int\limits_{\rm disc}
\Sigma(\bfR^\prime)\frac{\bfR^\prime}{R^{\prime 3}}\de^2 \bfR^\prime,
\label{eq:aid}
\end{align}
which is equal in magnitude and opposite to the acceleration exerted on the star by the planet and the disc, affects all objects in the system uniformly and enters both equations (\ref{eq:gasEoMRgen}) and (\ref{eq:plEoMRgen}) in the same way, alongside the relevant terms representing only the direct gravity. 

If one were to neglect the indirect forces entirely (i.e. drop $\aid$), the equations (\ref{eq:gasEoMRgen}) and (\ref{eq:plEoMRgen}) would become
\begin{align}
\ddot{\bf R}=&-GM_\star\frac{\bfR}{R^3}
-G\Mp\frac{\bfR-\bfRp}{\vert\bfR-\bfRp\vert^3}
\nonumber\\
&- G\int\limits_{\rm disc}
\Sigma(\bfR^\prime)
\frac{\bfR-\bfR^\prime}{\vert\bfR-\bfR^\prime\vert^3}\de^2 \bfR^\prime,
\label{eq:gasEoMRgen-noIT}\\
\ddot{\bf R}_\mathrm{p}  = & -GM_\star\frac{\bfRp}{\Rp^3}
- G\int\limits_{\rm disc}
\Sigma(\bfR^\prime)
\frac{\bfRp-\bfR^\prime}{\vert\bfRp -\bfR^\prime\vert^3}\de^2 \bfR^\prime.
\label{eq:plEoMRgen-noIT}
\end{align}
This reduction requires quantitative justification, which is not easy to provide in general. For example, if we consider the second and third terms proportional to $\Mp$ in the right-hand side of (\ref{eq:gasEoMRgen}), both of them have a general form $G\Mp f(\bfR,\bfRp)$, where $f$ is some function. Neglecting one of these terms and keeping the other may thus be inconsistent, unless we can explicitly assure that the magnitude of $f$ in one of the terms is small. As a result, in general one should keep (or neglect) both the direct and indirect terms due to a given mass, be it a planet (second and third terms in the right-hand side of (\ref{eq:gasEoMRgen})) or a disc (fourth and fifth terms proportional to $\Sigma$ in the right-hand side of (\ref{eq:gasEoMRgen})). Eliminating one of these contributions while keeping the other is not self-consistent, unless one can explicitly demonstrate the smallness of the neglected term. This is in line with the conclusions reached earlier in \citet{Crida2022,Crida2025}.

In this work we focus on studying a non-self-gravitating disc, which is often a good model of astrophysical discs. This means that when modelling disc dynamics, in equation (\ref{eq:gasEoMRgen}) the direct gravitational term due to other disc elements can be neglected, which is achieved by dropping the fourth term in the right-hand side of the equation (\ref{eq:gasEoMRgen}) responsible for the direct disc gravity. This is formally justified by taking a limit $\Sigma\to 0$. However, in that limit the (fifth) indirect term due to the disc in equation (\ref{eq:gasEoMRgen}) must vanish as well, implying that when the disc self-gravity is neglected, the indirect effect of the disc on its own dynamics must not be considered either \citep[cf.][]{Mittal2015}, in agreement with what we said earlier. 

This logic can be naturally extended in the opposite direction: when the indirect potential of the disc is dropped from the calculation (which again formally implies a limit $\Sigma\to 0$), ideally one should also not include disc self-gravity (unless one can explicitly demonstrate the smallness of the indirect contribution due to the disc). In other words, the direct gravitational and indirect forces of the disc (as well as any other massive component of the system, e.g. planets) should go hand in hand when exploring its dynamics \citep{Zhu2015}: they either need to be considered simultaneously or dropped altogether, which is an important statement of mathematical self-consistency.  

To summarize this reasoning, in the limit of a non-self-gravitating disc equation (\ref{eq:gasEoMRgen}) reduces to 
\ba
\ddot{\bf R}=-GM_\star\frac{\bfR}{R^3}
-G\Mp \left(\frac{\bfR-\bfRp}{\vert\bfR-\bfRp\vert^3}+\frac{\bfRp}{R_\mathrm{p}^3}\right),
\label{eq:gasEoMR-nosg}
\ea
which is the form we will be using in this work when looking at the disc dynamics in Sections \ref{sec:pert} \& \ref{sec:IT-AM}.

Note, however, that when studying the impact of disc-planet coupling on the planetary orbit in a non-self-gravitating disc (see Section \ref{sec:mig}), we cannot neglect the terms proportional to $\Sigma$ in the second line of equation (\ref{eq:plEoMRgen}), unlike in equation (\ref{eq:gasEoMRgen}). Dropping this term would imply neglecting the disc-planet coupling altogether, which is not what we want. As a result, for analyzing planetary motion in a non-self-gravitating disc we will be using the full equation (\ref{eq:plEoMRgen}), whereas for studying the impact of disc-planet coupling on the disc we will use the equation (\ref{eq:gasEoMR-nosg}).


\subsection{Physical setup}
\label{sec:physical}


To model gas dynamics of the disc, we employ an adiabatic, non-barotropic equation of state (EoS) with $\gamma = 1.4$, i.e. we allow for radially-varying entropy profile but do not employ additional heating or cooling terms. This thermodynamic setup preserves angular momentum of density waves propagating through the disc in the absence of nonlinear dissipation \citep[see][]{Miranda2020I}, simpifying our analysis in Section \ref{sec:AMF}. The disc is initially axisymmetric with surface density and temperature (expressed through the sound speed $\cs$) profiles in the power law form: 
\begin{align}
\Sigma_0(R) = \Sigma(\Rp) (R/\Rp)^{-p},~~~~~~\cs^2(R) = \cs^2 (\Rp) (R/\Rp)^{-q},
\label{eq:PLs}
\end{align}
where we set $p=q=1$. The initial disc aspect ratio is $h = H/R = \cs/(R \Omega) = 0.1$ and is radially constant for the chosen $q=1$.  The aspect ratio does not change as a result of shock heating in the course of our simulations because of the short duration of our runs (see below). 

We consider planetary masses of $\Mp = (0.01,0.3)\Mth$, where $\Mth=M_\star h^3$ is the characteristic thermal mass delineating the regimes of linear (for $\Mp\lesssim \Mth$) versus nonlinear  (for $\Mp\gtrsim \Mth$) density waves driven by the planet. As we will see later, nonlinear effects are small initially for $\Mp = 0.01\Mth$ but they become pronounced already for $\Mp = 0.3\Mth$, thus our choice of $\Mp$ allows us to probe different physical regimes. The planetary potential is softened to avoid singularities on the grid. We use a particular form of the softened potential given by the equation (24) of \citet{Cimerman2021}, with the softening length set at $r_s=0.6H$. 

The planetary orbit at $\Rp$ is held fixed over the course of our simulations. We infer the impact of the indirect forces on the planetary orbit by calculating the torque acting on it (see Section \ref{sec:mig}), which is ultimately based on the equations (\ref{eq:plEoMRgen}) or (\ref{eq:plEoMRgen-noIT}).


\subsection{Numerical simulations}
\label{sec:numerical}


To quantify the effect of the indirect term in disc-planet interaction we often resort to fully nonlinear hydrodynamic simulations. Most of them are carried out with {\sc athena++} \citep{Athenapp2020} in the non-inertial astrocentric (centred on the star), two-dimensional (2D) cylindrical-polar frame. The general setup of our models is similar to that of \citet{Cimerman2024b}, with the important difference that now we perform runs both including and excluding the indirect term (hereafter IT) in the hydrodynamic equation of motion for the disc fluid, by which we mean the last term in the equation (\ref{eq:gasEoMR-nosg}). 

Our numerical setup assumes a cylindrical grid spaced logarithmically in radius, in most cases ranging from $R=0.1 \Rp$ to $10 \Rp$ (sometimes we use more radially extended domains, out to 20$\Rp$) and uniformly in azimuth, spanning $0 \leq \phi \leq 2\pi$, using $N_R = 2640$ and $N_\phi = 3598$ cells, respectively. We note here that we increased the outer radius (doubled) compared to the models of \citet{Cimerman2024b} because in the analysis of our models we found that even with careful fine-tuning of the wave-damping zones (located at $0.2\Rp$ and $9\Rp$), we could not eliminate all reflections from the outer disc boundary. By running our simulations for a relatively short time of $t_\mathrm{end} \simeq 20 \Pp$, where $\Pp$ is the planetary orbital period (roughly the sound-crossing time over the domain of interest, here $R \leq 5\Rp$), this extra space in the outer disc acts as a buffer for the wave to propagate into before it can be reflected. 

To provide a useful point of comparison, we also run several simulations of the same system in the barycentric, inertial frame using {\sc disco}\footnote{
Specifically, we used the version \url{https://github.com/ajdittmann/Disco}.} \citep{2016ApJS..226....2D}. The domains of these simulations extended from $R=0.1 \Rp$ to $R=20 \Rp$, with $N_r=2304$ cells and a grid spacing that smoothly transitioned between linear to logarithmic at $R=0.2\Rp$. The number of azimuthal zones varied so as to keep $R\Delta\phi\approx\Delta R$ as closely as possible, resulting in $N_\phi\approx2031$. We employed wave damping zones a $R\leq0.2\Rp$ and $R\geq17\Rp$.


\subsection{Linear calculations}
\label{sec:linear}


To better understand some aspects of disc-planet coupling with or without the indirect term, we also used linear calculations \citep{KP93,RP12,PR12} performed using a linear framework developed earlier in \citet{Miranda2019I},  \citet{Fairbairn2022,Fairbairn2025a}. This framework solves numerically a set of linearized (and Fourier decomposed in azimuth) fluid equations for the disc perturbation (see Section \ref{sec:pert-lin} for some background). The full details of this method can be found in the aforementioned studies.

This framework allows us to provide a detailed analysis of disc perturbation in the limit of very low mass planets, $\Mp\ll\Mth$, which is performed in Section \ref{sec:pert}. It also naturally allows us to study important integral characteristics of the planet-driven density waves, such as disc torques and angular momentum flux (see Section \ref{sec:IT-AM}). This method provides a very powerful analysis tool, cross-checked against direct simulations in a number of disc-planet related problems: impact of non-ideal disc thermodynamics on density wave propagation \citep{Miranda2019I,Miranda2020I,Miranda2020II}, spawning of multiple arms by planet-driven density waves \citep{Miranda2019II}, negative torque density phenomenon \citep{RP12}, torque wiggles \citep{Cimerman2024b}, migration and eccentricity evolution of eccentric planets \citep{Fairbairn2025a,Fairbairn2025}, etc. Our linear calculations use the same physical setup (described in Section \ref{sec:physical}) as our simulations (including the adiabatic EoS, which is implemented via a $\beta$-cooling model with a very large value of $\beta$) allowing a direct comparison between them.


\section{Impact of indirect potential on the surface density perturbation}
\label{sec:pert}


We start by exploring the effect of the indirect force due to a planet on the structure of the planet-induced perturbation in the disc by comparing surface density perturbation patterns computed with and without the IT  and identifying the differences between them. Before moving on to the simulation results, we highlight some important features of indirect forces in the context of disc-planet coupling following from the linear theory \citep{GT80}. 

For a planet on a circular orbit, one can adopt a Fourier ansatz $\delta x(r,\phi,t) = \delta x_m(r) \exp[\mathrm{i}m(\phi-\Omega_\mathrm{p} t)]$ for any perturbed disc variable $\delta x$, where $\Omega_\mathrm{p}$ is the pattern frequency of the planetary perturbation (equal to its orbital Keplerian frequency) and $m$ is the azimuthal wavenumber. Linearizing the 2D hydrodynamic equations one obtains a very general equation for the evolution of the enthalpy-like variable $\delta h_m=\delta P_m/\Sigma$ ($P$ and $\Sigma$ being the 2D pressure and surface density) in the form \citep{Miranda2020I}
\ba
\label{eq:master}
\frac{\mathrm{d}^2}{\mathrm{d}R^2}\delta h_m + C_1\frac{\mathrm{d}}{\mathrm{d}R}\delta h_m + C_0\delta h_m = \Psi_m,
\ea
where $C_1$ and $C_0$ are functions of the radial distance $R$, with their explicit form being set by the background disc properties (e.g. radial profiles of $\Sigma_0$, $\cs$, etc.) and thermodynamic assumptions. The right hand side $\Psi_m$ depends on disc properties and thermodynamic assumptions, as well as on the $m$-th Fourier component of the perturbing planetary potential $\Phi_m$ and its derivatives. More specific versions\footnote{See also \citet{Fairbairn2022} for an extension to the case of an eccentric planetary orbit.} of the equation  (\ref{eq:master}) have been derived by \citet{GT79} for a barotropic (constant entropy) EoS, by \citet{Baruteau2008}, \citet{Tsang2014} for a non-barotropic, adiabatic (radially varying entropy) EoS, by \citet{Lee2016} for a locally isothermal EoS (in the homogeneous form), and by \citet{Miranda2020I} for a disc with $\beta$-cooling. The key point here is that regardless of disc thermodynamics, in linear theory $\Phi_m$ affects only the $m$-th component of the disc perturbation. 

The planetary indirect term $-G\Mp\bfRp/\Rp^3$, i.e. the last term in equation (\ref{eq:gasEoMR-nosg}), is the gradient of the planetary indirect potential 
\begin{align}
\Phi_\mathrm{id}(\bfR)=G\Mp\frac{\bfRp\cdot\bfR}{\Rp^3}
=G\Mp \frac{R}{\Rp^2}\cos\left(\phi-\Omega_\mathrm{p} t\right),
\label{eq:id_pot1}
\end{align}
where the last equality assumes planet on a circular orbit
at an azimuthal angle $\phi=\Omega_\mathrm{p} t$. This form implies that in a linear limit the inclusion of the IT should affect only the $m=1$ component of the perturbation $\delta h_1$ when compared to the calculation without the IT, with $\delta h_m$ not changing for $m\neq 1$.

\citet{GT80} showed that a perturbation component $\delta h_m$ gets excited by $\Phi_m$ at the inner/outer Lindblad resonances where the angular frequency of the disc is $\Omega=\Omegap m/(m\pm 1)$ (lower/upper sign). In the radial range between the resonances, perturbations are evanescent (exponentially damped) but propagate as a wave outside this region. In a Keplerian disc, the $m=1$ inner Lindblad resonance is absent (it formally resides at $R\to 0$), while the outer Lindblad resonance is located at $R_1=2^{2/3}\Rp\approx 1.59\Rp$. Based on the above description, one expects the $m=1$ component of the perturbation associated with the IT to propagate as a wave outside $R_1$ and be exponentially suppressed for $R<R_1$. 

We now illustrate these expectations (and deviations from them) with simulations in both linear and nonlinear regimes.


\subsection{Low-$\Mp$ (linear) regime}
\label{sec:pert-lin}


We start by examining the low $\Mp=0.01\Mth$ case, when the disc-planet interaction is in the linear regime (as $\Mp\lesssim \Mth$) and the linear theory outlined above should work well.

\begin{figure}
	\begin{center}
	\includegraphics[width=0.49\textwidth]{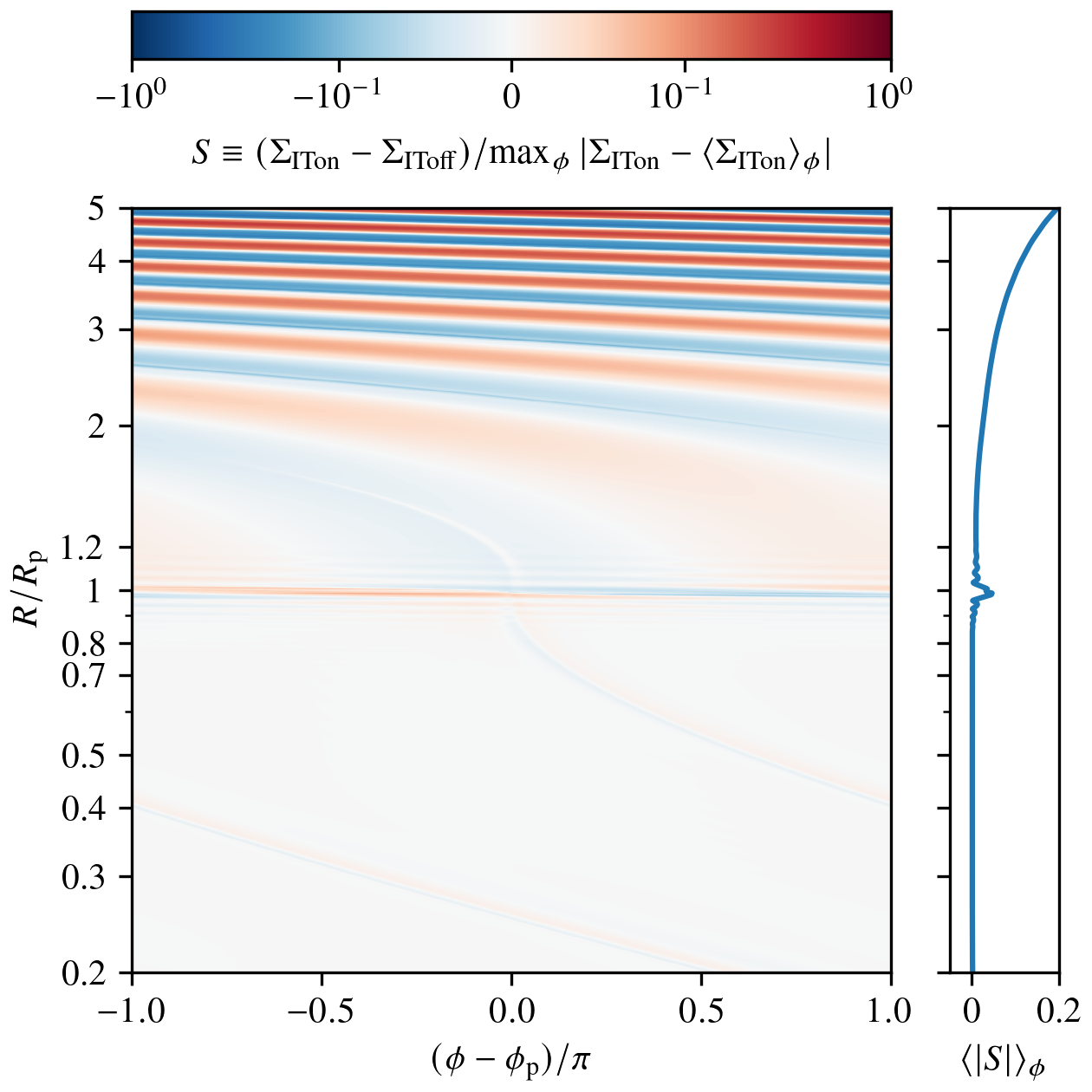}
	\caption{
2D map of the difference $S$ of the surface densities $\Sigma(R,\phi)$ from simulations with/without the indirect term, normalized by the local amplitude of non-axisymmetric perturbation (i.e of density wave) in a simulation with IT. Both runs use $\Mp = 0.01 \Mth$ and all other parameters are identical. Right panel is the radial profile of the azimuthal average of $|S|$ providing an idea of the relative role played by the IT.}
	\label{fig:2d_diff_sigma_IT_no_IT_001}
	\end{center}
\end{figure}

In Figure \ref{fig:2d_diff_sigma_IT_no_IT_001} we show the map of the relative difference $S$ in surface densities established in two disc-planet simulations with the IT included ($\Sigma_\mathrm{ITon}$) and without it ($\Sigma_\mathrm{IToff}$), but otherwise with identical parameters. The difference $\Sigma_\mathrm{ITon}-\Sigma_\mathrm{IToff}$ is normalized at each $R$ by the maximum of the amplitude of the non-axisymmetric component of $\Sigma$ (i.e. planet-induced density wave) in a simulation with the IT included, max$_\phi|\Sigma_\mathrm{ITon}-\langle\Sigma_\mathrm{ITon}\rangle_\phi|$. Normalizing the $\Sigma$ difference in such a way results in $S$ being independent of $\Mp$ and allows one to see how the modification of $\Sigma$ caused by the exclusion of IT compares with the amplitude of the density wave itself. 

Looking at Figure \ref{fig:2d_diff_sigma_IT_no_IT_001} one immediately notices that the $\Sigma$ difference has an $m=1$ morphology at every $R\gtrsim \Rp$, which is precisely what the linear theory predicts. This difference has a wave-like pattern, which get tightly-wound as $R$ increases, consistent with the linear dispersion relation for density waves in differentially rotating discs \citep{OL02}. 

The amplitude of the $\Sigma$ difference is very different in the inner and outer discs, as further illustrated by the right hand panel where we show the azimuthal average of $|S|$ at each $R$. This provides an idea of how large the $\Sigma$ differences induced by IT are, compared to the overall density wave amplitude. One can see that $S$ is extremely small in the inner disc, which is consistent with the expectation of exponentially damped $m=1$ linear perturbation in that region. However, in the outer disc, $S$ has a substantial amplitude increasing with $R$, which is not surprising given that the forcing potential $\Phi_\mathrm{id}$ grows with $R$. By $R=5\Rp$ the difference $|\Sigma_\mathrm{ITon}-\Sigma_\mathrm{IToff}|$ is about $20\%$ of the local wave amplitude, meaning that dropping the IT from the calculation results in a misrepresentation of the density wave (i.e. perturbation pattern) characteristics at that level.   

\begin{figure}
	\begin{center}
	\includegraphics[width=0.49\textwidth]{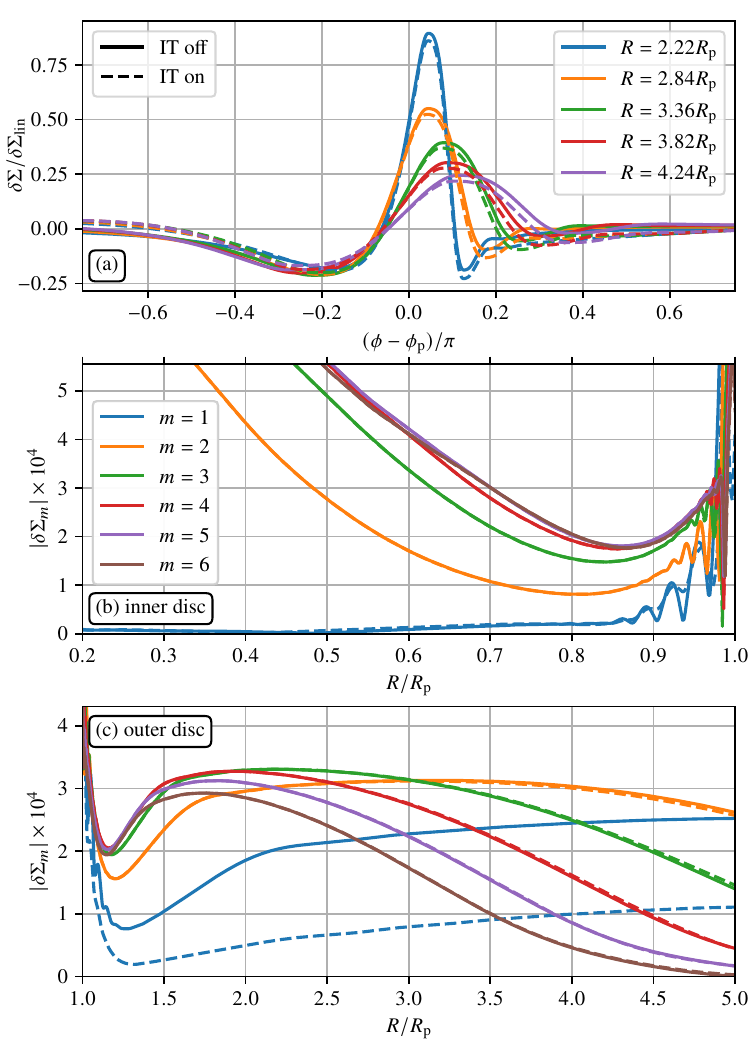}
	\caption{
        Comparison of the disc response to the planetary perturbation computed for a low planetary mass $\Mp = 0.01 \Mth$, corresponding to the linear regime of disc-planet interaction. Simulations are run for an adiabatic disc with $\gamma=1.4$, $\hp = 0.1$, temperature and surface density power law indices $p=q=1$.
		(a) Comparison of the azimuthal profiles of the surface density perturbation $\delta\Sigma$ of the planet-driven density wave at different radii $R$ in the outer disc (normalized by the linear theory prediction $\delta\Sigma_\mathrm{lin}$, see Section \ref{sec:analytical}) extracted from simulations with (dashed) and without (solid) the indirect term. Only small differences are visible. (b),(c) Azimuthal Fourier components of $\delta\Sigma$ for several low values of $m$ in the inner (b) and outer (c) regions of the disc, obtained from simulations with (dashed) and without (solid) the indirect term. One can see that indirect term affects only $m=1$ component of $\delta\Sigma$ and predominantly in the outer disc, as expected from linear theory, which applies in this case because of small $\Mp$.   }
	\label{fig:fourier_comparison}
	\end{center}
\end{figure}

\begin{figure}
	\begin{center}
	\includegraphics[width=0.49\textwidth]{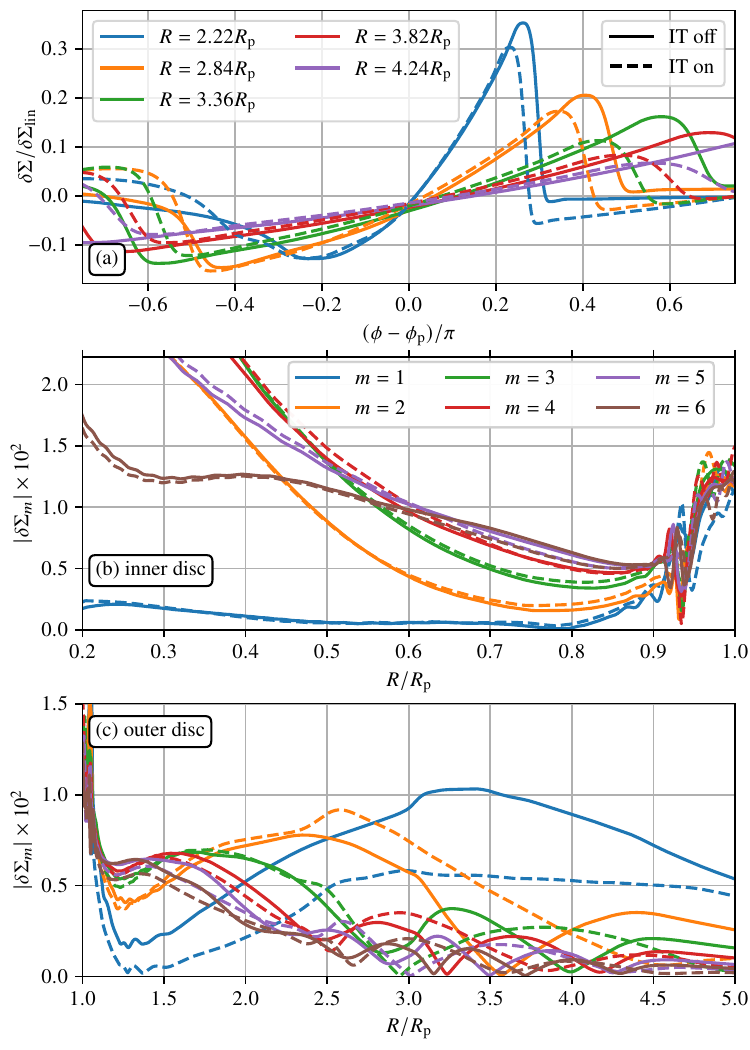}
	\caption{
Same as Fig. \ref{fig:fourier_comparison} but for $\Mp = 0.3 \Mth$. One can see that the differences between the calculations with the IT and without it are considerably larger than in Fig. \ref{fig:fourier_comparison}, no longer confined only to $m=1$ perturbation component, and noticeable also in the inner disc. The oscillatory structure of $\delta\Sigma_m$ in the outer disc is explained in Appendix \ref{sec:fourier_oscillations}.
  }
	\label{fig:fourier_comparison_0.3}
	\end{center}
\end{figure}

A more in-depth analysis of these differences is provided in Figure \ref{fig:fourier_comparison}. In panel (a) we show the azimuthal profiles of the surface density perturbation $\delta\Sigma=\Sigma-\langle\Sigma\rangle_\phi$ near the density wave maximum at several different radii in the outer disc. We display $\delta\Sigma$ from the runs with the IT (dashed) and without it (solid), normalized by the linear theory prediction for wave amplitude $\delta\Sigma_\mathrm{lin}$ \citep[see][]{R02,Cimerman2021}. One can see that the density wake profile slowly evolves due to the nonlinear effects present even at such low $\Mp$, decaying in amplitude and broadening azimuthally, and the differences between $\delta\Sigma$ computed with and without the IT are rather small in real space but increase with $R$. 

Next, in panels (b) and (c) we plot the amplitudes of azimuthal Fourier harmonics of $\delta\Sigma$, $\delta\Sigma_m$, for $m=1-6$ as functions of $R$ inside and outside\footnote{Rapid increase of $|\delta\Sigma_m|$ near $\Rp$ is caused by the quasi-static atmosphere accumulating around the planet and should be ignored.} $\Rp$, again with and without the IT. In the inner disc, one can see that $|\delta\Sigma_m|$ with or without the IT are the same --- dashed lines fall on top of the solid ones for all $m\ge 2$. The same is true also for the $m=1$ component, but the more important thing is that $|\delta\Sigma_1|$ is very small in amplitude and rapidly decays towards the star. This is very different from $|\delta\Sigma_m|$ for $m\ge 2$, which grow as $R$ decreases as a result of approximate conservation of the angular momentum flux \citep{R02}. This difference in behaviour is caused by the absence of the $m=1$ inner Lindblad resonance in a Keplerian disc and the lack of direct $m=1$ mode excitation in the inner disc, unlike for the higher $m$ modes. The only reason why $|\delta\Sigma_1|$ is non-zero but small for $R<\Rp$ is due to the exponentially damped (due to tunneling through the corotation region) $m=1$ perturbation excited at the outer Lindblad resonance at $R_1$.

In the outer disc (panel (c)) we see a similar picture for all $m\ge 2$ components $|\delta\Sigma_m|$ --- they are essentially the same with and without the IT, although here $|\delta\Sigma_m|$ decay with the distance from the planet as a result of nonlinear evolution, which is faster for higher $m$. However, there is a dramatic difference between $|\delta\Sigma_1|$ computed with and without the IT, with the former having a significantly lower magnitude than the latter. All this is entirely consistent with the linear expectation that the effect of the IT should be present only in the $m=1$ component of the perturbation.

Overall, low-$\Mp$ simulations fully support linear theory predictions outlined earlier.


\subsection{High-$\Mp$ (nonlinear) regime}
\label{sec:pert-nonlin}


Next we examine the higher $\Mp=0.3\Mth$ case, when the disc-planet coupling becomes more nonlinear. In Figure \ref{fig:fourier_comparison_0.3} we show the results of a calculation for this higher $\Mp$, presented analogous to the one in Figure \ref{fig:fourier_comparison}. 

In this case, panel (a) shows that there is now a noticeable difference between the density wake shapes in real space obtained with and without the IT; this difference grows with the distance traveled by the wave in the outer disc. Panel (b) shows that in the inner disc the deviations between the calculations with and without the IT, while small, become noticeable for all $m$. The power in the $m=1$ mode in the inner disc is still very low compared to $m\ge 2$ modes, as expected from linear theory. However, it is noticeably non-zero and slowly increases as $R\to 0$, in contradiction with the linear expectations of pure exponential damping. Our tests demonstrate that this slowly increasing inwards $m=1$ mode is caused by the emergence of non-zero eccentricity in the inner disc for $\Mp=0.3\Mth$, possibly driven by the mechanisms outlined in \citet{Teys2016}; see also \citet{Crida2025} and the discussion of Figure \ref{fig:dTidR} in Appendix \ref{sec:torque-sep}. Disc eccentricity is negligible in the $\Mp=0.01\Mth$ run.  

However, the most dramatic difference from the linear expectation is revealed by panel (c) of Figure \ref{fig:fourier_comparison_0.3}, which shows that $|\delta\Sigma_m|$ computed with and without the IT are now considerably different from each other for all $m$ and not just for $m=1$. This propagation of the differences in $|\delta\Sigma_m|$ to $m>1$ harmonics of the perturbation must result from nonlinear effects that couple different $m$-modes, allowing the transfer of some power from $m=1$ to higher-$m$ harmonics. This process eventually feeds the differences between the dashed and solid curves in Figure \ref{fig:fourier_comparison_0.3}c for all $m$. 

We note that these differences are not due to some secondary nonlinear structures, e.g. vortices, that may appear in simulations at high $\Mp$ and drive their own density waves. First, our simulations are run for a very short time, $20\Pp$, while the vortices take longer to form due to the Rossby Wave Instability \citep{Cimerman2023}. Second, vortices would drive density waves not only in the outer regions but also in the inner disc. However, differences between the solid and dashed curves are small in the inner disc. Third, the deviations between the dashed and solid curves in Figure \ref{fig:fourier_comparison_0.3}c are larger for lower $m$, as expected from a gradual leakage of power from $m=1$ mode to higher $m$ due to wave nonlinearity. 

One curious aspect of the Figure \ref{fig:fourier_comparison_0.3}c is that the decay of $|\delta\Sigma_m|$ in the outer disc is no longer monotonic with $R$. Instead, $|\delta\Sigma_m|$ curves show oscillatory behavior, going to zero at certain locations, with the frequency of these radial oscillations being higher for higher $m$. In Appendix \ref{sec:fourier_oscillations} we provide an explanation for this behavior and show that it is caused by the {\it azimuthal spreading} of the wake profile caused by its nonlinear evolution \citep{GR01,R02,Dong2011b,Duffell2012,Cimerman2021}, a process which is much faster in the $\Mp=0.3\Mth$ case (Fig. \ref{fig:fourier_comparison_0.3}a) than in the $\Mp=0.01\Mth$ run (Fig. \ref{fig:fourier_comparison}a).

Regardless of this curious phenomenon, the main message of Figure \ref{fig:fourier_comparison_0.3} is that even in the mildly nonlinear case (and certainly for higher $\Mp$) accounting for the IT leads to differences with the calculation neglecting the IT, and these differences spread across multiple Fourier modes of $\delta\Sigma$, in contrast to the linear case.


\section{Torque on the disc and wave angular momentum}
\label{sec:IT-AM}


Our next goal is to understand the impact of indirect forces on the angular momentum exchange in disc-planet coupling, starting with the disc; the effect on the dynamics of the planet will be explored in Section \ref{sec:mig}. As alluded to in Section \ref{sec:intro}, this exchange is a complicated multi-step process that is mediated by the excitation, propagation and dissipation of planet-driven density waves. Each individual step of this coupling can be characterized by a certain angular momentum-based characteristic. At the stage of density wave excitation, the planetary potential injects angular momentum into the wave at a rate (per unit radius and time) given by the {\it excitation} torque density $\de T(R)/\de R$. The wave accumulates this angular momentum and carries it through the disc, with the angular momentum flux $F_J$ characterizing the amount of wave angular momentum crossing a given radius $R$ per unit of time. As the wave subsequently dissipates, it transfers its angular momentum to the disc at the rate (per unit radius and time) given by the {\it deposition} torque density $\de T_\mathrm{dep}(R)/\de R$. We now systematically examine the impact of the IT on each of these angular momentum characteristics.


\subsection{Excitation torque density}
\label{sec:torque-disc}


We start by computing the excitation torque density in 
a frame comoving with the star. The specific (per unit disc mass) torque ${\bf t}= \bfR\times\ddot\bfR$ in astrocentric coordinates acting on a fluid element of a non-self-gravitating disc can be written using equation (\ref{eq:gasEoMR-nosg}) as
\begin{align}
{\bf t} =
G\Mp\left(\bfR\times\bfRp\right)\left(\frac{1}{\vert\bfR-\bfRp\vert^3}-\frac{1}{\Rp^3}\right)={\bf t}_\mathrm{d}+{\bf t}_\mathrm{id},
\label{eq:TR-nonsg}
\end{align}
where we split ${\bf t}$ into the direct ${\bf t}_\mathrm{d}$ and indirect ${\bf t}_\mathrm{id}$ contributions:
\begin{align}
{\bf t}_\mathrm{d}(\bfR) &=G\Mp\frac{\bfR\times\bfRp}{\vert\bfR-\bfRp\vert^3},~~~~~~
{\bf t}_\mathrm{id}(\bfR)=-G\Mp\frac{\bfR\times\bfRp}{\Rp^3}.
\label{eq:splits}
\end{align}
Obviously, the stellar potential does not contribute to ${\bf t}$. In a 2D disc all these torques have components only along $z$-axis. 

The (radial) excitation torque density $\de T/\de R$ --- the amount of torque (its $z$-component) exerted by a planet on a radial annulus of the disc --- can then be expressed as
\begin{align}
\frac{\de T(R)}{\de R} = R\int_0^{2\pi}\Sigma(R,\phi)\, t_z(R,\phi)\, \de\phi
 =\frac{\de T_\mathrm{d}(R)}{\de R}+\frac{\de T_\mathrm{id}(R)}{\de R},
\label{eq:dTdR}
\end{align}
where $t_z={\bf e}_z\cdot{\bf t}$ is the $z$-component of ${\bf t}$. In the last expression we separated $\de T(R)/\de R$ into the {\it direct} torque density due to the direct gravitational force of the planet 
\begin{align}
\frac{\de T_\mathrm{d}(R)}{\de R} = G\Mp R\int_0^{2\pi}\Sigma(R,\phi)\frac{\left(\bfR\times\bfRp\right)_z}{\vert\bfR-\bfRp\vert^3}\,\de\phi,
\label{eq:dTdRd}
\end{align}
and the {\it indirect} part of the astrocentric torque density due to the IT: 
\begin{align}
\frac{\de T_\mathrm{id}(R)}{\de R} = -\frac{G\Mp}{\Rp^3}\,R\int_0^{2\pi}\Sigma(R,\phi) \left(\bfR\times\bfRp\right)_z\de\phi.
\label{eq:dTdRid}
\end{align}
Note that the direct torque density could be obtained by using the non-self-gravitating version of equation (\ref{eq:plEoMRgen-noIT}), instead of (\ref{eq:gasEoMR-nosg}), in equation (\ref{eq:dTdR}) in order to obtain ${\bf t}$, but then the indirect part would be totally missed, as expected. Also, one can set $\Sigma\to\delta\Sigma$ in equations (\ref{eq:dTdRd}) and (\ref{eq:dTdRid}) since the planet exerts no torque on the axisymmetric component of $\Sigma$. 

To the best of our knowledge, all existing studies of disc-planet coupling that computed excitation torque density due to a planet did this by dropping $\de T_\mathrm{id}/\de R$ entirely and accounting only for $\de T_\mathrm{d}/\de R$ (see also Section \ref{sec:lit}). It is important to keep in mind that the IT enters this calculation in two ways: first, through the calculation of the perturbation $\delta\Sigma(R,\phi)$ as described in Section \ref{sec:pert} and, second, explicitly through $\de T_\mathrm{id}/\de R$. But even studies that properly include the IT when computing perturbed $\Sigma$, still consistently neglect the $\de T_\mathrm{id}/\de R$ contribution to the excitation torque density. It should be clear from our derivation of equation (\ref{eq:dTdR}) that such a neglect must result in a non-conservation of the angular momentum involved in disc-planet coupling in the astrocentric frame. And indeed, we demonstrate next that dropping $\de T_\mathrm{id}/\de R$ leads to rather dramatic consequences.

\begin{figure}
	\begin{center}
	\includegraphics[width=0.49\textwidth]{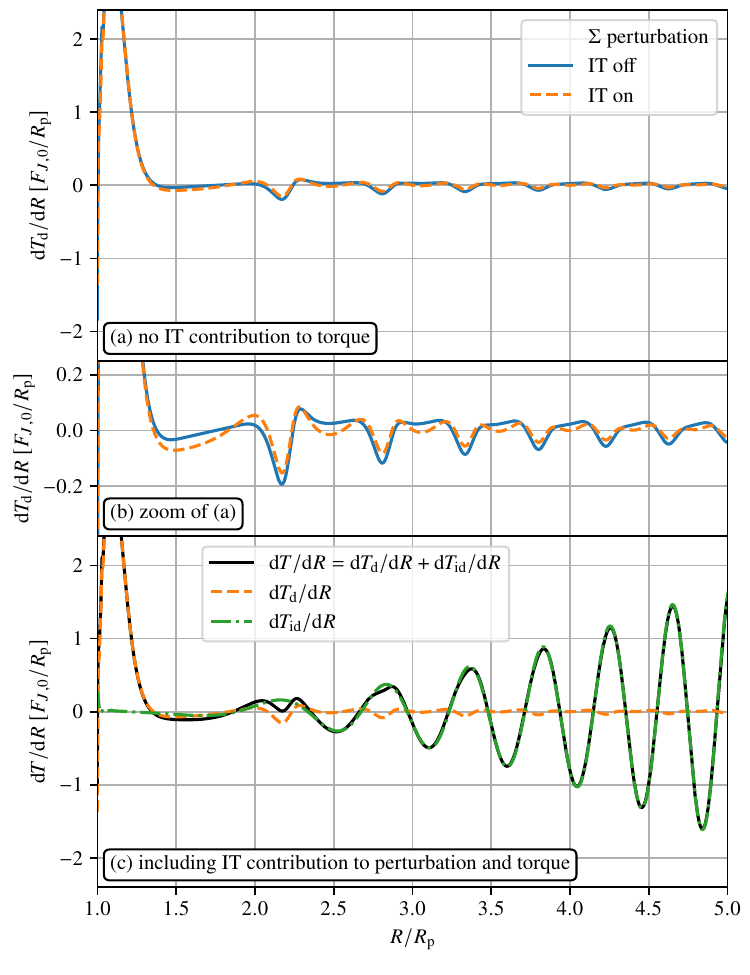}
	\caption{
		Illustration of the torque density computed (a,b) without the IT and (c) with the IT included. Data for the outer disc ($R>\Rp$) are taken from the {\sc athena++} simulations for an adiabatic disc with $\hp = 0.1$, $p=q=1$ and $\Mp = 0.01 \Mth$. In panel (a), only the direct torque density $\de T_\mathrm{d}/\de R$ given by the equation (\ref{eq:dTdRd}) is shown; the solid blue curve is $\de T_\mathrm{d}/\de R$ computed based on $\Sigma$ from a simulation without the IT, while the dashed orange curve is the direct torque computed using $\Sigma$ from a simulation with the IT properly included. Panel (b) is a vertical zoom of panel (a) illustrating the torque wiggles \citep{Cimerman2024b}. In panel (c) the $\Sigma$ data are from the simulation with the IT included, and the different curves show the full torque $\de T/\de R$ (black solid) defined by equation (\ref{eq:dTdR}), the direct torque $\de T_\mathrm{d}/\de R$ (orange dashed, same curve as in panels (a),(b)), and the indirect torque $\de T_\mathrm{id}/\de R$ (green dot-dashed). Note the conspicuous oscillations of growing amplitude exhibited by the indirect and total torque densities as $R$ increases in panel (c).
  }
	\label{fig:dTdR}
	\end{center}
\end{figure}

\begin{figure}
	\begin{center}
	\includegraphics[width=0.49\textwidth]{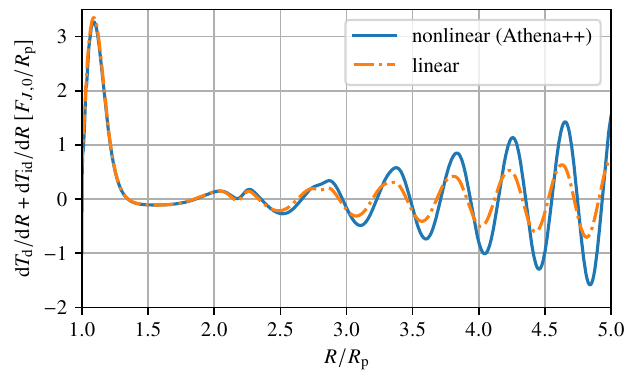}
	\caption{
        Comparison of the full excitation torque density $\de T/\de R$, including the indirect contribution $\de T_\mathrm{id}/\de R$, obtained in our $\Mp=0.01\Mth$ simulation with {\sc athena++} (solid blue) and computed using linear calculation (dot-dashed orange). Both calculations produce oscillating $\de T/\de R$; the difference in amplitudes of oscillations is discussed in Section \ref{sec:analytical}.
  }
	\label{fig:dTdR_IT_lin_vs_nonlin}
	\end{center}
\end{figure}

In Figure \ref{fig:dTdR} we show the results of torque calculations performed on the outputs of two $\Mp=0.01\Mth$ simulations --- including and excluding the IT in the calculation of the perturbed $\Sigma$. In the former case we also have a freedom of including or excluding $\de T_\mathrm{id}/\de R$ in the calculation of the excitation torque density. Torque densities are normalized by the natural reference unit in disc-planet interaction, $F_{J,0}/\Rp$, where $F_{J,0}$ is the characteristic one-sided torque \citep{GT80,GR01}
\begin{align}
    F_{J,0} = \left(\frac{\Mp}{M_\star}\right)^2 \hp^{-3} \Sigma_\mathrm{p} \Rp^4 \Omegap^2=
    \Mp\frac{\Mp}{\Mth} \frac{\Sigma_\mathrm{p} \Rp^2}{M_\star} \Rp^2\Omegap^2.
    \label{eq:FJ0}
\end{align}

In panel (a) we show in detail the direct torque density $\de T_\mathrm{d}/\de R$ computed in the outer disc ($R>\Rp$) via equation (\ref{eq:dTdRd}) for both runs. The solid blue curve uses $\Sigma$ from a simulation in which the IT term was absent (`IT off'), while for the dashed orange curve $\Sigma$ comes from a simulation with the IT term included (`IT on'). One can see that the two curves closely follow each other around the peak of $\de T_\mathrm{d}/\de R$ near the planet, while further out one notices low-amplitude, quasi-periodic, decaying features --- the torque wiggles previously studied in \citet{Cimerman2024b}. In panel (b) we zoom in vertically to better illustrate the torque wiggles and also to highlight how the different ways of computing $\Sigma$ (with and without the IT) impact $\de T_\mathrm{d}/\de R$. One can see that the two curves show small differences reflecting the differences in the perturbation patterns produced by the two simulations, as described in Section \ref{sec:pert}. However, both of them eventually decay to zero as the distance from the planet increases. So far, the picture is consistent with other existing calculations of the excitation torque density \citep[e.g][]{DAngelo2008,Arza18,Cimerman2024b}.

The story changes dramatically in panel (c), where we show $\de T/\de R$ (solid black) incorporating both the direct (dashed orange) as well as the indirect (dot-dashed green) components of the torque density. This calculation uses $\delta\Sigma$ from a run fully accounting for the IT. The direct part $\de T_\mathrm{d}/\de R$ is the same as the dashed curves in panels (a), (b) and decays with $R$. But the indirect contribution $\de T_\mathrm{id}/\de R$ shows a very different behavior: it oscillates and the amplitude of oscillations grows with $R$. As a result, far from the planet in the outer disc $\de T/\de R$ is entirely dominated by the indirect contribution $\de T_\mathrm{id}/\de R$ and also increases with $R$ without a limit. This behaviour is a dramatic change from any existing disc torque calculation. 

To ensure that these torque oscillations are not an artefact of the nonlinear calculation, in Figure \ref{fig:dTdR_IT_lin_vs_nonlin} we show $\de T/\de R$ (dot-dashed orange curve) obtained using a linear calculation (see Section \ref{sec:linear}) with the IT fully included in calculations of both $\Sigma$ and $\de T/\de R$ and scaled to $\Mp=0.01\Mth$ used in simulations. For comparison, the solid blue curve is the simulation-based $\de T/\de R$, i.e. same as the black curve in Figure \ref{fig:dTdR}c. One can see that the linear calculation reproduces growing oscillations of the torque density with the same periodicity as the fully nonlinear simulations. The difference in the amplitudes of oscillations will be addressed later in Section \ref{sec:analytical}. But the clear presence of growing $\de T/\de R$ oscillations in both linear and nonlinear calculations implies that they are a real physical effect.   

We now show that the radially-diverging behavior of the indirect torque $\de T_\mathrm{id}/\de R$ arises because of the reflex motion of the central star. Let us a consider a mass (e.g. a fluid element in the disc) in the barycentric frame of a star-planet system with a finite specific angular momentum ${\bf l}_\mathrm{b}={\bf r}\times \dot{\bf r}$, where ${\bf r}$ is the barycentric position of this mass. In the astrocentric coordinates the position of this mass is $\bfR={\bf r}-{\bf r}_\star$, where ${\bf r}_\star$ is the stellar position in the barycentric frame. Then the specific angular momentum ${\bf l}_\mathrm{a}$ of that mass in the astrocentric frame is 
\ba
{\bf l}_\mathrm{a}={\bf R}\times \dot{\bf R}={\bf l}_\mathrm{b}-{\bf r}\times \dot{\bf r}_\star-{\bf r}_\star\times \dot{\bf r}+{\bf r}_\star\times \dot{\bf r}_\star.
\label{eq:lsplit}
\ea
Now consider what happens as we increase the distance of the mass from the central star, when $r,R\to\infty$. If the mass is bound and has a finite $|{\bf l}_\mathrm{b}|$, its velocity $|\dot {\bf r}|\to 0$. Since $|{\bf r}_\star|$ is finite, the third term in the right hand side of (\ref{eq:lsplit}) vanishes, while the first and last terms remain finite. However, the second term in  (\ref{eq:lsplit}) clearly diverges in the limit of $r\to\infty$ (because of the infinitely large lever arm and finite velocity $\dot{\bf r}_\star$ of stellar reflex motion), meaning that $|{\bf l}_\mathrm{a}|$ also diverges despite $|{\bf l}_\mathrm{b}|$ being finite. 

Moreover, because of planet-induced stellar motion around the barycenter of the system, $\dot{\bf r}_\star$ varies in time and so does ${\bf l}_\mathrm{a}$, with an infinitely growing amplitude as $r\to\infty$. This implies that in the astrocentric frame an infinitely large torque must be applied to that distant mass, thereby explaining the divergence of the indirect torque for $r,R\to \infty$ as a consequence of stellar reflex motion. These conclusions are corroborated further in Section \ref{sec:torque}. 

The oscillatory behaviour of $\de T/\de R$ in the outer disc may still seem puzzling and raises natural questions regarding the angular momentum conservation in the disc. It calls for deeper analysis to understand its origin, which is what we do in the rest of this section.


\subsection{Wave AMF and integrated torque}
\label{sec:AMF}


Torque exerted by a planet on the disc, as described in the previous Section, contributes to the angular momentum of the planet-driven density wave \citep{GT79,GT80}. The wave carries this angular momentum away from the planet, a process which is characterized by another key angular momentum characteristic --- the wave angular momentum flux (AMF), $F_J(R)$. In the astrocentric frame, AMF represents the amount of angular momentum carried by the wave across a cylinder of radius $R$ centered on the star per unit of time. Mathematically, wave AMF can be expressed as
\ba
F_J(R)=R^2\int_0^{2\pi}\Sigma v_R\delta v_\phi\,\de\phi,
\label{eq:FJ}
\ea
where $\delta v_\phi$ is the deviation of $v_\phi$ from its azimuthally-averaged value, and all velocities are defined relative to the star. 

Wave angular momentum conservation \citep[enforced by our thermodynamic assumptions, see][]{Miranda2020I} dictates that in the absence of explicit dissipation, the radial gradient of $F_J(R)$ (i.e. the amount of angular momentum added to the wave per unit radius and time) should be equal to the torque density $\de T(R)/\de R$ at the same radius \citep[e.g.][]{Cordwell2024},
\ba
\frac{\de F_J(R)}{\de R}=\frac{\de T(R)}{\de R}.
\label{eq:Tsplit1}
\ea
Indeed, according to \citet{Gold1989}, in an ideal (non-dissipative) fluid perturbed by a rigidly rotating external potential, no secular changes in the angular momenta of fluid elements are possible (except at corotation). Therefore, the disc does not change its state and all angular momentum injected by the external potential gets transported away by internal stresses, i.e. waves. Another way to phrase this and interpret equation (\ref{eq:Tsplit1}) is to say that $F_J(R)$ should match the full (integrated) torque $T(R)$ exerted by the planet on the disc between $\Rp$ and $R$,
\ba  
T(R)=\int_{\Rp}^R
\frac{\de T(R^\prime)}{\de R}\de R^\prime.
\label{eq:Tdef}
\ea
For that reason, we will be comparing $F_J$ with $T$ and its direct and indirect components $T_\mathrm{d}$ and $T_\mathrm{id}$, defined using $\de T_\mathrm{d}/\de R$ and $\de T_\mathrm{id}/\de R$, respectively, as the integrand in equation (\ref{eq:Tdef}), with $T=T_\mathrm{d}+T_\mathrm{id}$.

\begin{figure}
	\begin{center}
	\includegraphics[width=0.49\textwidth]{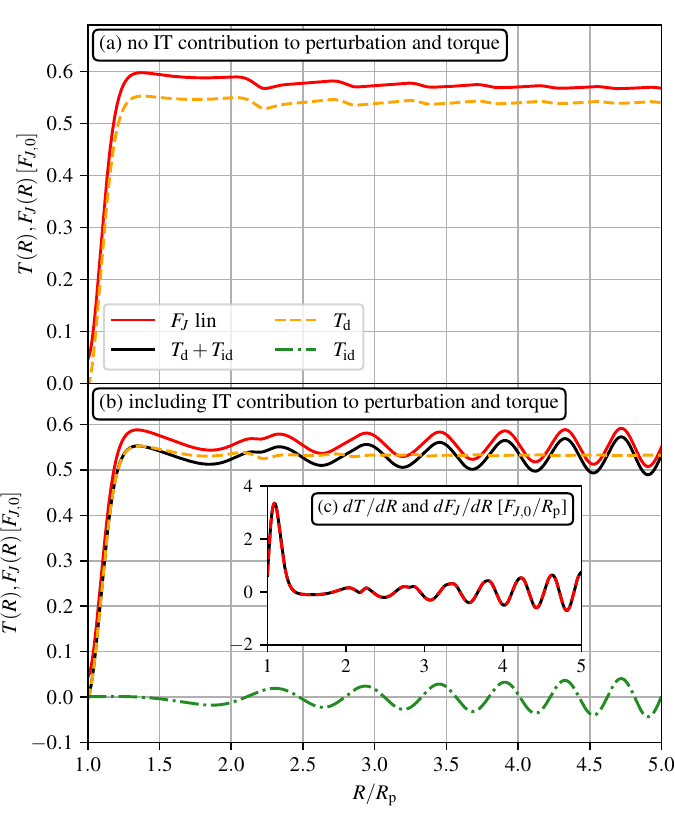}
	\caption{
    Radial profiles of the angular momentum flux $F_J$ of the planet-driven density wave (solid red), and integrated direct torque $T_\mathrm{d}$ (dashed orange), indirect torque $T_\mathrm{id}$ (dot-dashed green), and total torque $T$ (solid black). These are linear adiabatic calculations (see Section \ref{sec:linear}) with no wave dissipation (unlike in Figure \ref{fig:AMF}). In panel (a) IT is fully neglected, $T_\mathrm{id}=0$, and $T=T_\mathrm{d}$. In panel (b) both the $\Sigma$ perturbation and the torque calculation account for the IT. Except for the small vertical offset (due to the corotation effects), $T(R)$ and $F_J(R)$ curves follow each other, which is also illustrated in the inset (panel (c), same colors scheme as in (b) but with the red curve dashed), showing that their radial derivatives are the same, see Eq. (\ref{eq:Tsplit1}). In panel (a), the low-amplitude features of  decaying amplitude in $T$ and $F_J$ are the torque wiggles \citep{Cimerman2024b}, whereas the oscillations of increasing amplitude in panel (b) are due to the indirect contribution to the torque. See text for details. 
  }
	\label{fig:AM-lin}
	\end{center}
\end{figure}

To set the stage, we start with a linear calculation, akin to Section \ref{sec:linear}. In Figure \ref{fig:AM-lin} we show the AMF and integrated torques obtained through the linear framework and normalized by $F_{J,0}$. In panel (a) we illustrate a calculation that neglects the IT altogether, both in the computation of the perturbed $\Sigma$ and in the torque evaluation, in which case $T_\mathrm{id}=0$ and $T=T_\mathrm{d}$. The solid red curve shows the AMF $F_J(R)$ while the dashed orange curve is $T_\mathrm{d}(R)=T(R)$. The two curves closely follow each other with just a small constant vertical offset, which can be attributed to the corotation torque effects. Both curves show a rapid rise near the planet, where $\de T/\de R$ peaks, but then flatten out with small, decaying oscillations on top --- the torque wiggles \citep{Cimerman2024b} manifesting themselves in both the integrated torque and AMF. This is a standard behaviour previously seen in all calculations ignoring the IT \citep[e.g.][]{Dong2011,RP12}. 

The situation is quite different in panel (b) of Figure \ref{fig:AM-lin}, where we illustrate a calculation fully accounting for the IT, including the computation of $\Sigma$. We again show the AMF (solid red) and integrated direct torque $T_\mathrm{d}$ (dashed orange), but now also the integrated indirect torque $T_\mathrm{id}$ (dot-dashed green) and the full integrated torque $T$ (solid black). The behaviour of $T_\mathrm{d}$ is very similar to what one sees in panel (a), even despite the somewhat different $\Sigma$ structure. But now there is also a non-zero $T_\mathrm{id}$, and it shows the familiar oscillations in the outer disc, as expected from the behavior of $\de T_\mathrm{id}/\de R$ in Figure \ref{fig:dTdR}c. It is natural that these oscillations are then also reflected\footnote{Just for reference, in the inner disc $T_\mathrm{id}(R)$ is very small, so that $T(R)$ and $T_\mathrm{d}(R)$ show only minor differences.} in the profile of $T(R)$. We verified that the amplitude of these oscillations increases with $R$ without limit. 

What is most interesting though is that AMF, $F_J$, also exhibits radial oscillations, despite it being calculated independently from the velocity components, i.e. in a completely different way than the integrated torques. As a result, $F_J$ tracks $T$ very well, with only a small vertical offset. This is a clear manifestation of the wave angular momentum conservation in the astrocentric frame, additionally emphasized in an inset, panel (c), where we demonstrate that the equation (\ref{eq:Tsplit1}) is satisfied exactly. Thus, for our linear calculation, which has no dissipation by design, all torque exerted by the planet indeed goes into the angular momentum of the density wave.

\begin{figure}
	\begin{center}
	\includegraphics[width=0.49\textwidth]{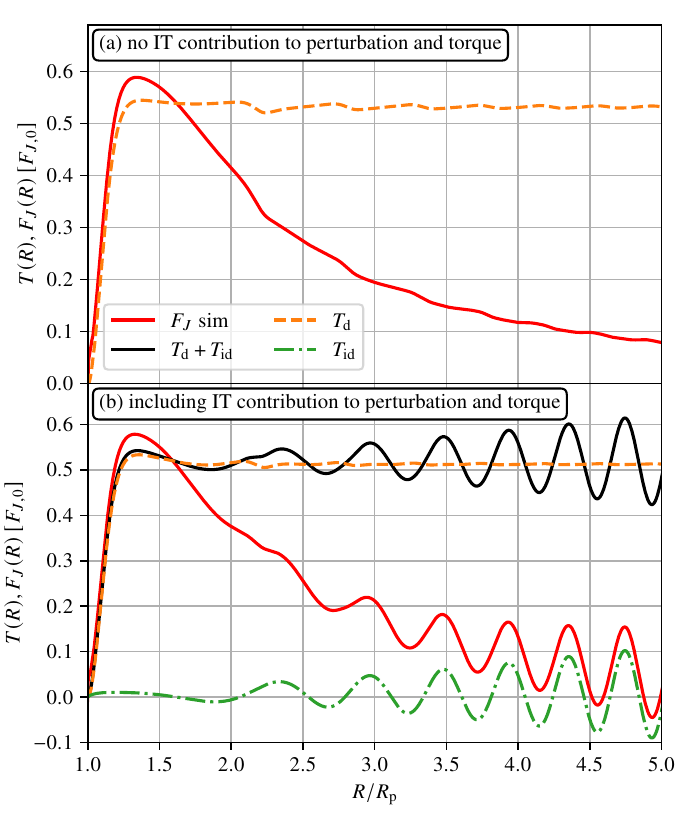}
	\caption{
    Similar to Figure \ref{fig:AM-lin}, but now based on simulations with $\Mp=0.01\Mth$, without (a) and with (b) the IT; meaning of the curves is the same. A notable difference is the decay of $F_J$ with $R$, caused by the nonlinear dissipation of the planet-driven density wave. See text for details. 
  }
	\label{fig:AMF}
	\end{center}
\end{figure}

These linear calculations are further corroborated by the results of our nonlinear simulations in Figure \ref{fig:AMF}, where we show $F_J(R)$ and integrated torques from two simulations in the astrocentric frame with a $\Mp=0.01\Mth$ planet. In panel (a) we again do not account for the IT both when computing the disc structure and when computing the torque, so that $T(R)=T_\mathrm{d}(R)$. One can see that near the planet $F_J(R)$ and $T(R)=R_\mathrm{d}(R)$ curves again follow each other closely (up to a constant vertical offset), which is a consequence of equation (\ref{eq:Tsplit1}) since the wave does not dissipate initially. But then around $1.4\Rp$ the density wave shocks \citep{GR01,R02} and starts to dissipate. Its AMF $F_J$ starts to decay since wave angular momentum gets transferred to the disc fluid (see Section \ref{sec:depos}). As a result, the $F_J$ curve falls below $T_\mathrm{d}$ and then steadily decays as $R$ increases; a careful inspection shows the signs of weak torque wiggles \citep{Cimerman2024b} imprinted on both the $T(R)$ and $F_J(R)$ curves. This picture is in agreement with a number of earlier studies \citep{Dong2011b,Miranda2019II,Cimerman2021,Cimerman2024b} that did not include the IT.

In panel (b) of Figure \ref{fig:AMF} the IT is properly included in the calculation of both the disc $\Sigma$ and $T(R)$. One immediately notices that the $F_J$ curve looks strikingly different compared to panel (a): as $R$ increases, $F_J$ starts exhibiting oscillations of a growing amplitude superimposed on the shock-induced decay of $F_J$. These oscillations are so substantial that $F_J$ starts reaching negative values at $R>4.5\Rp$. The $T(R)$ curve exhibits analogous oscillations, leading us to the natural conclusion that the oscillations of $F_J(R)$ reflect the radial pattern of the applied torque even in the presence of wave dissipation. And, just as in Figures \ref{fig:dTdR} and \ref{fig:AM-lin}, the origin of these oscillations can be traced back to the IT as the $T_\mathrm{id}(R)$ curve shows them as well, while $T_\mathrm{d}(R)$ looks similar to panel (a). To the best of our knowledge, such divergent oscillations of $F_J(R)$ have not been explicitly reported for the existing disc-planet simulations carried out in the non-inertial frame and including the IT (see Section \ref{sec:lit} for a detailed comparison with the existing literature).  

We can now re-assess the issue of angular momentum conservation when including the IT, as highlighted at the end of Section \ref{sec:torque-disc}. What Figures \ref{fig:AM-lin} and \ref{fig:AMF} show is that accounting for the IT in torque calculations is not an impediment to the angular momentum conservation, but is in fact an integral part. Indeed, if we set $T_\mathrm{id}=0$ and $T=T_\mathrm{d}$ but kept the IT in the calculation of $\delta\Sigma$, the AMF would still exhibit prominent oscillations in the outer disc. But now, these oscillations would be unmatched by the torque acting on the disc, meaning that the angular momentum of the density wave would not be conserved. Thus, conservation of the wave angular momentum in the astrocentric frame actually {\it requires} one to properly include $T_\mathrm{id}$ when the IT is active in the disc structure calculation. 

\begin{figure}
	\begin{center}
	\includegraphics[width=0.49\textwidth]{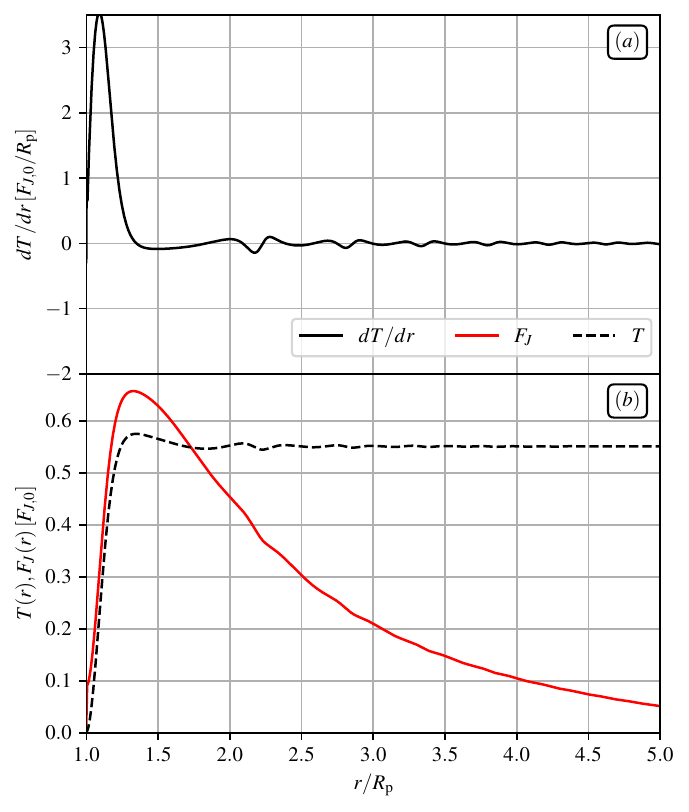}
	\caption{
    Radial profiles of the (a) torque density $\de T/\de r$ and (b) integrated torque $T(r)$ (black dashed) and angular momentum flux $F_J(r)$ of the planet-driven density wave as functions of the barycentric distance $r$. These profiles are derived from our barycentric simulations performed with {\sc disco} and using the same parameters as Figure \ref{fig:AMF}. All curves show evidence of low-amplitude torque wiggles \citep{Cimerman2024b} but no diverging oscillations as in Figure \ref{fig:AMF}b.
  }
	\label{fig:disco}
	\end{center}
\end{figure}

We should also emphasize that the concept of $T_\mathrm{id}$ and the divergent oscillatory behaviors of $T$ and $F_J$ are unique to the non-inertial astrocentric frame. In the inertial, e.g. barycentric, frame indirect forces are absent, obviating $T_\mathrm{id}$ altogether and changing the behavior of $T$ and $F_J$. We illustrate this using a simulation with the same parameters as in Figure \ref{fig:AMF} but carried out in the barycentric frame using {\sc disco}, see Section \ref{sec:numerical}. In Figure \ref{fig:disco} we show the torque density (panel (a)), integrated torque $T$ and angular momentum flux $F_J$ (panel (b)) as functions of the barycentric distance $r$, which is not very different from $R$ since $\Mp\ll M_\star$. The torque-related quantities are computed including torques exerted by both the planet and the star (both of which are displaced from the barycentric frame's origin), and $F_J$ uses the disc velocity components relative to the system's barycenter. One can see that, as expected, all these metrics show a regular behavior as $r$ increases: $\de T/\de r\to 0$, $T(r)$ saturates at a constant level, and $F_J(r)$ steadily decays as a result of nonlinear wave dissipation. While these curves exhibit the decaying torque wiggles, they do not show the divergent oscillations at large $r$, as typical for the non-inertial frame. Their behavior is qualitatively reminiscent of Figures \ref{fig:dTdR}a and \ref{fig:AMF}a, in which only the direct gravitational force was considered, although the perturbation structure and coordinate systems are different.


\subsection{Deposition torque density}
\label{sec:depos}


The decay of AMF $F_J$ relative to integrated torque $T$ in Figure \ref{fig:AMF} at large $R$ is caused by the dissipation of the density wave. Our simulations do not have any explicit sources of wave dissipation (e.g. viscosity or radiation), and numerical dissipation is very small, so that wave damping is driven predominantly by nonlinear effects --- steepening of the wave profile leading to its shocking and dissipation \citep{GR01,R02}. This mechanism is effective even for low $\Mp$, when the nonlinearity is weak. But in a purely linear calculation this process and the associated wave dissipation are absent by default, as Figure \ref{fig:AM-lin} demonstrates.

Angular momentum lost by the density wave as a result of its dissipation gets transferred to the disc fluid. The disc gains angular momentum at the rate (per unit $R$ and time) given by the deposition torque density 
\ba
\frac{\de T_\mathrm{dep}}{\de R}=\frac{\de T}{\de R}-\frac{\de F_J}{\de R},
\label{eq:Tdep}
\ea
as follows from the wave angular momentum conservation. Non-zero $\de T_\mathrm{dep}/\de R$ is what ultimately drives disc evolution, leading to phenomena such as gap formation \citep{R02b,Kanagawa2015,Cordwell2024}. Thus, knowing $\de T_\mathrm{dep}/\de R$ is crucial for understanding evolution of planet-induced disc structures. This motivates us to examine the effect of the IT on the deposition torque density.  

\begin{figure}
    \centering
    \includegraphics[width=\columnwidth]{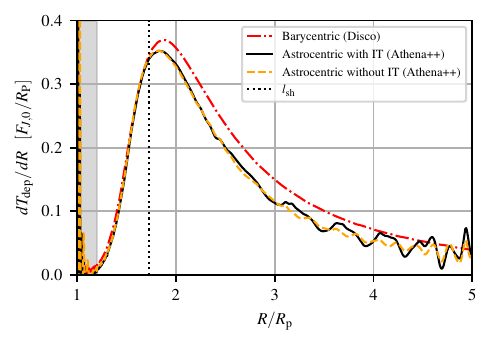}
    \caption{Deposition torque density $\de T_\mathrm{dep}/\de R$ in the outer disc derived from  {\sc athena++} simulations with (solid black) and without (dashed orange) the IT. Dot-dashed red curve shows $\de T_\mathrm{dep}/\de R$ obtained in a barycentric simulation with the same parameters performed with {\sc disco}. The vertical dotted line marks the nominal shock position. The gray band around $R=\Rp$ is a corotation region. See Section \ref{sec:depos} for details. 
    }
    \label{fig:bary_astro_deposition}
\end{figure}

In Figure \ref{fig:bary_astro_deposition} we plot $\de T_\mathrm{dep}/\de R$ derived from simulations with the IT included (solid black), based on Figure \ref{fig:AMF}b, and without the IT (dashed orange), based on Figure \ref{fig:AMF}a. One can see that the differences between the two curves are rather minimal, meaning that the inclusion or exclusion of the IT when computing $\Sigma$ and the torque has a very weak effect on $\de T_\mathrm{dep}/\de R$. The prominent diverging oscillations of both $\de T_\mathrm{id}/\de R$ and $\de F_J/\de R$ standing out in the outer disc in a calculation with the IT fully included (see Figure 7), effectively cancel each other out in the calculation of $\de T_\mathrm{dep}/\de R$, leaving a well behaved, decaying deposition torque density in Figure \ref{fig:bary_astro_deposition}, just like in the case without the IT. Small irregular oscillations of $\de T_\mathrm{dep}/\de R$ visible in the outer disc represent a numerical artefact leading to imperfect cancellation of torque wiggles in $F_J(R)$ and $T(R)$. The nominal shock position $l_\mathrm{sh}$ seems to correlate better with the peak of $\de T_\mathrm{dep}/\de R$ rather than its initial rise. However, in higher resolution runs the peak shifts to the right and $\de T_\mathrm{dep}/\de R$ is suppressed for $|R-\Rp|\lesssim l_\mathrm{sh}$, as expected. 

Note that if we did not account for the indirect torque $\de T_\mathrm{id}/\de R$ in the torque density calculation, then $\de T_\mathrm{dep}/\de R$ curve would feature the divergent oscillations in the outer disc inherited from $\de F_J/\de R$, incompatible with the actual disc evolution. This observation additionally emphasizes the importance of properly including the indirect torque in the density wave angular momentum budget. 

We also show $\de T_\mathrm{dep}/\de R$ obtained in a very different way, namely from a  {\sc disco} simulation with the same parameters but carried out in the barycentric frame, see Figure \ref{fig:disco}. As indirect forces are absent by default in this frame, this calculation provides an independent check of the effect of the damping density wave on the disc, derived from our astrocentric simulations. The result is shown by the red dot-dashed curve (as a function of barycentric distance $r$, which should not make much difference), and matches the overall shape of $\de T_\mathrm{dep}/\de R$ quite well. There are some differences, e.g. $\de T_\mathrm{dep}/\de R$ in the barycentric frame is slightly larger than in the astrocentric frame, but nevertheless the agreement is quite good given the significant differences between the numerical methods used in this comparison. 

The key message from Figure \ref{fig:bary_astro_deposition} is that at least for low mass planets the IT plays a rather minor role in the behavior of $\de T_\mathrm{dep}/\de R$, meaning that planet-induced disc evolution should also be rather insensitive to the inclusion or exclusion of the IT in simulations.


\subsection{Oscillatory behaviour of the indirect torque}
\label{sec:analytical}


Having analyzed the behaviour of wave AMF and torques in the non-inertial frame, we now provide a mathematical understanding of the oscillatory character of the indirect torque density $\de T_\mathrm{id}(R)/\de R$ in application to disc-planet interaction. In Appendix \ref{sec:torque-deriv} we derive an analytical expansion for the full torque density $\de T/\de R$ in the astrocentric frame. In particular, the indirect component of the torque density is found to be 
\begin{align}
\frac{\de T_\mathrm{id}(R)}{\de R}  & =2\pi G \Mp\left(\frac{R}{\Rp}\right)^2\,\IM\,\Sigma_1(R),
\label{eq:Tid}
\end{align}
where $\Sigma_1(R)$ is the $m=1$ coefficient of the Fourier expansion of $\Sigma(R,\phi)$ defined as
\ba
\Sigma_m(R)=(2\pi)^{-1}\int_0^{2\pi}\Sigma(R,\phi)\,\mathrm{e}^{-\I m\phi}\,\de\phi.
\label{eq:Sig_m}
\ea  
Since the unperturbed disc is axisymmetric, $\Sigma_1$ is equal to the $m=1$ harmonic of the density perturbation $\delta\Sigma_1$.

In the following we will limit ourselves to considering low-mass $\Mp\lesssim\Mth$ planets on circular orbits. In this case, the planet-induced $\delta\Sigma$ is (i) in the linear regime ($\delta\Sigma\lesssim\Sigma$) and (ii) stationary in the frame co-rotating with the planet, meaning that $\Sigma(R,\phi,t)=\Sigma(R,\phi)$ as long as $\phi$ is measured relative to the instantaneous $\bfRp$ direction (the origin of $\phi$ does not affect the torque acting on an annulus of the disc). Then $\delta\Sigma_1$ is excited at the outer $m=1$ Lindblad resonance located at $R_1=2^{2/3}\Rp$ (see Section \ref{sec:pert}). In the inner disc $\delta\Sigma_1$ is evanescent and its amplitude is very small, so that $\de T_\mathrm{id}/\de R$ is non-oscillatory and small there. 

In the outer disc, far outside $R_1$ one can use the tight-winding approximation and write
\begin{align}
\delta\Sigma_1(R)=\delta\tilde\Sigma_1(R)\exp\left(\mathrm{i}\int^R_{R_1} k_R(R^\prime)\de R^\prime\right),
\label{eq:Sig_anzatz}
\end{align}
where $\delta\tilde\Sigma_1$ is the complex amplitude of $\delta\Sigma_1$ (Arg$\,\delta\tilde\Sigma_1$ is constant with $R$) and $k_R$ is the WKB radial wavenumber in a Keplerian disc given by \citep{GT79} 
\begin{align}
	k^2_{R} = \cs^{-2}
		\left[ 
			m^2 (\Omega - \Omegap)^2 - \Omega^2
		\right]=\frac{\Omegap}{\cs^2}\left(\Omegap-2\Omega\right), 
	\label{eq:k_WKB}
\end{align}
where we set $m=1$ to obtain the last equality. Also notice that $k_R(R_1)=0$, as expected. Substituting this into the expression (\ref{eq:Tid}) one obtains
\begin{align}
\frac{\de T_\mathrm{id}(R)}{\de R}  & =2\pi G \Mp
\,
\vert\delta\tilde\Sigma_1(R)\vert\frac{R^2}{\Rp^2}
\nonumber\\
&\times \sin\left[\mathrm{Arg}\,\delta\tilde\Sigma_1-\int_{R_1}^R\cs^{-1}\sqrt{\Omegap\left(\Omegap-2\Omega\right)}\de R^\prime\right].
\label{eq:Tid1}
\end{align}

This expression provides an explicit illustration of the oscillatory nature of the indirect torque: as $R$ increases, so does the integral in the argument of $\sin$ in equation (\ref{eq:Tid1}), resulting in radial oscillations of $\de T_\mathrm{id}/\de R$. The radial distance between the successive nodes (nulls) $R_k$ of $\de T_\mathrm{id}(R)/\de R$
in the outer disc is given (implicitly) by the condition
\begin{align}
\int_{R_\mathrm{k}}^{R_\mathrm{k+1}} \cs^{-1}(R^\prime)\sqrt{\Omegap\left[\Omegap-2\Omega(R^\prime)\right]} \de R^\prime=\pi. 
\label{eq:Deltaphi}
\end{align}
Far from the planet, where $\Omega\ll\Omegap$, one finds
\begin{align}
R_\mathrm{k+1}-R_\mathrm{k}\approx\pi \,\Omegap^{-1}c_s(R_\mathrm{k})=\pi\,H_\mathrm{p}\left(R_k/\Rp\right)^{-q/2}, 
\label{eq:Deltaphi-as}
\end{align}
where $H_\mathrm{p}$ is the disc scale height at the planetary location. For a disc illustrated in Figure \ref{fig:dTdR}, $q=1$ and the distance between the nodes decreases as $R_k^{-1/2}$, which is noticeable in the behavior of the green dot-dashed curve in panel (c) of that figure. Equations (\ref{eq:Deltaphi})-(\ref{eq:Deltaphi-as}) are essentially analogous to the periodicity conditions formulated for the torque wiggles in disc-planet coupling \citep{Cimerman2024b} and for the radial oscillations of the gravitational torque density in circumbinary discs \citep[see Eqs. (17), (18) of][]{Cimerman2024a} since all of these phenomena depend on the global spatial structure of the planet-driven density wave.

Furthermore, the dependence of $\de T_\mathrm{id}(R)/\de R$ on only the $m=1$ component of $\delta\Sigma$ allows us to explain the intriguing discrepancy between the linear and simulation-based amplitudes of the $\de T/\de R$ oscillations seen in Figure \ref{fig:dTdR_IT_lin_vs_nonlin}. Naively, one might expect the simulation-based $\de T_\mathrm{id}(R)/\de R$ oscillations (solid) to have lower amplitude than the one predicted by the linear theory (dot-dashed), since in simulations the density wave amplitude decays as $R$ increases due to shock dissipation (see Figure \ref{fig:fourier_comparison}a). However, Figure \ref{fig:fourier_comparison}c shows that $\delta\Sigma_1$ actually {\it increases} with $R$, unlike all of the other $\delta\Sigma_m$ components. This phenomenon is caused by the nonlinear broadening of the wake as it propagates through the disc, which effectively transfers some power into the $m=1$ mode of the perturbation while the modes with $m>1$ get gradually extinguished by shock damping. This process has been previously described in \citet{Cimerman2024b} and is illustrated by their Figure 14. As a result, far from the planet the simulation-based $\delta\Sigma_1$ exceeds the linear prediction for $\delta\Sigma_1$, explaining the trends in $\de T/\de R$ behavior observed in Figure \ref{fig:dTdR_IT_lin_vs_nonlin}.


\subsection{Divergence of the indirect torque as $R\to\infty$}
\label{sec:torque}


We now analyze more carefully the origin of the divergence of the indirect torque density $\de T_\mathrm{id}/\de R$ as $R\to \infty$, which is clearly visible in Figure \ref{fig:dTdR}; this analysis reinforces simple arguments offered in the end of Section \ref{sec:torque-disc}. In principle, the mathematical reason for this divergence is already obvious from equation (\ref{eq:dTdRid}): according to its definition, $\de T_\mathrm{id}/\de R$ is proportional to $R^2$, which drives the divergence in the first place. However, the actual asymptotic behavior of $\de T_\mathrm{id}/\de R$ also depends on the behavior of $\delta\Sigma(R)$, which is something that we can address now.

Equation (\ref{eq:Tid1}) shows that the amplitude of $\de T_\mathrm{id}(R)/\de R$ oscillations depends on $R$ as $\propto R^2|\delta\tilde\Sigma_1(R)|$, i.e. it increases with $R$ as long as $|\delta\tilde\Sigma_1(R)|$ decays slower than $R^{-2}$. This is clearly the case in the examples shown in Figures \ref{fig:dTdR}c, \ref{fig:AM-lin}b, \ref{fig:AMF}b, since the oscillation amplitude grows in a divergent fashion as $R\to \infty$. In the linear regime, in the absence of dissipation, one can show \citep{R02,Miranda2020I} that far in the outer disc, where $\Omega\ll\Omegap$,
\ba
|\delta\tilde\Sigma_m(R)| \propto \left(\frac{\Sigma \Omegap}{R\cs^3}\right)^{1/2} ,
\label{eq:lin-amp}
\ea
for all $m$. Equations (\ref{eq:Tid1}) and (\ref{eq:lin-amp}) then imply that in a disc described by equation (\ref{eq:PLs}) the amplitude of $\de T_\mathrm{id}(R)/\de R$ oscillations should scale $\propto R^{(3/2)+(3q/4)-(p/2)}$. Thus, unless the surface density profile is very steep (i.e. $p$ is very high) the amplitude of $\de T_\mathrm{id}(R)/\de R$ will diverge as $R\to \infty$. 

After the wave shocks (which happens earlier for higher $\Mp$), the nonlinear wave dissipation causes $|\delta\tilde\Sigma_1(R)|$ to fall below the linear prediction (\ref{eq:lin-amp}). However, this has a rather limited effect in slowing down the growth of $\de T_\mathrm{id}(R)/\de R$ oscillations\footnote{As a result of nonlinear damping, the wave AMF would eventually decay as $F_J\propto t^{1/2}$, where the function $t(R)$ is given by the equation (32) of \citet{R02}. One can show that in this regime the amplitude of $\de T_\mathrm{id}(R)/\de R$ oscillations would scale as $\propto R^{(11/8)+(7q/16)-(5p/8)}$, not very far from the linear case (\ref{eq:lin-amp}).}. Moreover, as Figure \ref{fig:fourier_comparison}c illustrates,  in low-$\Mp$ cases $|\delta\Sigma_1|$ may actually increase with $R$ (due to the transfer of power to $m=1$ mode as discussed in the end of Section \ref{sec:analytical}) easily enabling the divergence of $\de T_\mathrm{id}(R)/\de R$ oscillations in Figure \ref{fig:dTdR}c.


\section{Torque on the planet and planetary orbital evolution}
\label{sec:mig}


Having studied the effect of the planetary IT on disc structure and angular momentum, we now explore how the IT due to the disc modifies the torque ${\bf T}_\mathrm{p}$ experienced by a planet gravitationally interacting with the disc. This torque is ultimately responsible for planet migration and eccentricity evolution in the course of disc-planet interaction. 

We start by emphasizing that in this Section the IT arises because of the reflex motion of the star caused by the gravity of the disc, and not of the planet as was the case in Sections \ref{sec:pert} and \ref{sec:IT-AM}. Taking a cross product of equation (\ref{eq:plEoMRgen}) with $\Mp\bfRp$ one finds that the torque on the planet due to the disc (its only non-zero $z$-component $T_\mathrm{p}$) is \citep{Fairbairn2025a}
\begin{align}
T_\mathrm{p}=
G\Mp\int\limits_{\rm disc}
\Sigma({\bf R})\left({\bf R}_\mathrm{p}\times{\bf R}\right)_z
\left(\frac{1}{\vert{\bf R}_\mathrm{p}-{\bf R}\vert^3}-\frac{1}{R^3}\right)\de^2{\bf R}\,,
\label{eq:Tp}
\end{align}
where we could again replace $\Sigma\to\delta\Sigma$ as only the non-axisymmetric part of $\Sigma$ leads to non-zero $T_\mathrm{p}$. 

As before, we split $T_\mathrm{p}=T_\mathrm{p,d}+ T_\mathrm{p,id}$ into the direct and indirect contributions. Comparing with equation (\ref{eq:dTdRd}), one can see that the direct torque on the planet defined as 
\begin{align}
T_\mathrm{p,d}=
G\Mp\int\limits_{\rm disc}
\Sigma({\bf R})
\frac{\left(\bfRp\times\bfR\right)_z}{\vert{\bf R}_\mathrm{p}-{\bf R}\vert^3}\,\de^2{\bf R}
=-\int\limits_\mathrm{disc}
\frac{\de T_\mathrm{d}(R^\prime)}{\de R}\,\de R^\prime
\label{eq:Tpd}
\end{align}
is equal in magnitude (and opposite in sign) to the direct torque exerted by the planet on the full disc. Thus, the total mutual {\it direct} astrocentric torque in the disc-planet system is zero. 

However, the same is not the case for the indirect disc torque exerted on the planet 
\begin{align}
T_\mathrm{p,id}=-G\Mp\int\limits_{\rm disc}
\Sigma({\bf R})
\frac{\left(\bfRp\times\bfR\right)_z}{R^3}\,\de^2{\bf R}.
\label{eq:Tpid}
\end{align}
According to equation (\ref{eq:dTdRid}), the full indirect planetary torque on the disc is
\begin{align}
\int\limits_\mathrm{disc}
\frac{\de T_\mathrm{id}(R^\prime)}{\de R}\de R^\prime=-\frac{G\Mp}{\Rp^3}\int\limits_\mathrm{disc}\Sigma(\bfR) \left(\bfR\times\bfRp\right)_z \de^2\bfR,
\label{eq:Tpid2}
\end{align}
which is clearly different from $T_\mathrm{p,id}$. Moreover, while $T_\mathrm{p,id}$ is finite since the integral in (\ref{eq:Tpid}) converges, the indirect torque on the disc (\ref{eq:Tpid2}) diverges as the outer extent of the disc expands (see Section \ref{sec:torque}). 

This asymmetry of indirect torques also implies that the {\it total} torque in the disc-planet system in the astrocentric frame is non-zero,
\begin{align}
T_\mathrm{p}+\int\limits_\mathrm{disc}
\frac{\de T(R^\prime)}{\de R}\de R^\prime\neq 0,
\label{eq:Tpdsum}
\end{align}
and in fact diverges as the disc size increases because of the behavior of $\de T/\de R$, with $T_\mathrm{p}$ being finite\footnote{$T_\mathrm{p}$ converges faster than the direct term $T_\mathrm{p,d}$ alone as the disc size increases since for $R\gg \Rp$
\ba
\frac{1}{\vert\bfRp-\bfR\vert^3}-\frac{1}{R^3} \approx 3\frac{\bfRp\cdot\bfR}{R^5}={\cal O}\left(R^{-4}\right)
\ea  
In this limit the IT suppresses the direct contribution of the distant masses to the integrand in (\ref{eq:Tp}). In other words, the effect of distant parts of the disc on the planet is tidal in its character.}. As a result, when the IT is properly included, one cannot compute the disc torque on the planet as just being the opposite of the planetary torque on the disc. 

Moreover, this also means that the full angular momentum of the disc-planet system is not conserved in the astrocentric frame. This result, which may seem paradoxical at first sight, is in fact fully in line with the findings of \citet{Rafikov2025}, who showed that the angular momentum of a system of gravitating objects --- the star, planet and disc, in our case --- in a non-inertial frame attached to one of them, is not conserved because of the torque exerted on the system by the indirect force.  

With rare exceptions, studies computing the disc torque on the planet accounted only for the direct contribution $T_\mathrm{p,d}$, neglecting the indirect part $T_\mathrm{p,id}$ altogether. Recently, \citet{Fairbairn2025a} explicitly considered $T_\mathrm{p,id}$ in their semi-analytical linear study of orbital evolution of low-mass eccentric planets and showed it to be small compared to $T_\mathrm{p,d}$. A similar conclusion was subsequently reached in \citet{Crida2025} based on simulations in the context of Type I migration, whereas for more massive (gap-opening) planets they found $T_\mathrm{p,id}$ to play a more significant role \citep[see also][]{RegalyVorobyov2017}. We now re-assess the relative contribution of the indirect component $T_\mathrm{p,id}$ of the disc torque on the planet in our setup using (semi-)analytical linear calculations. 


\subsection{Low mass planets}
\label{sec:id-contrib-p}


To evaluate the relative contribution of $T_\mathrm{p,id}$ to $T_\mathrm{p}$, we start by recalling that the one-sided direct planetary torques on the disc are $\sim F_{J,0}$ and similar in magnitude but opposite in sign. The full direct torque given by equation (\ref{eq:Tpd}) then ends up being \citep{Ward1997}
\begin{align}
T_\mathrm{p,d}\sim \hp F_{J,0}\propto \hp^{-2}\,,
\label{eq:Tpd_est}
\end{align}
as follows from equation (\ref{eq:FJ0}).

For the indirect part of disc torque acting on planet, setting $\phi$ to be the angle between $\bfRp$ and $\bfR$, we can rewrite (\ref{eq:Tpid}) as 
\begin{align}
T_\mathrm{p,id}& = -G\Mp\Rp\int_0^\infty
\frac{\de R}{R}\int_0^{2\pi}\Sigma(R,\phi)\sin\phi\,\de\phi
\nonumber\\
& = -2\pi G\Mp\Rp\int_0^\infty
\frac{\IM\,\Sigma_1(R)}{R}\,\de R,
\label{eq:Tpid1}
\end{align}
compare with equations (\ref{eq:Tid}) and (\ref{eq:Sig_m}). We note that for $R-R_1\gtrsim H_\mathrm{p}$, outside the $m=1$ OLR at $R_1$, $\Sigma_1=\delta\Sigma_1$ has a form given by the equations (\ref{eq:Sig_anzatz}) and (\ref{eq:k_WKB}). As a result, the integrand in (\ref{eq:Tpid1}) rapidly oscillates in this region, so that this part of the disc provides a small contribution to $T_\mathrm{p,id}$. On the other hand, for $R_1-R\gtrsim H_\mathrm{p}$, inwards of the $m=1$ OLR, $\delta\Sigma_1$ decays exponentially as $R$ decreases (see Figure \ref{fig:fourier_comparison}b), except at $R=\Rp$ where the perturbation is co-orbital with the planet and becomes large. As a result, one expects that the main contributions to the integral in equation (\ref{eq:Tpid1}), as well as in equation (\ref{eq:Tpid}), come from the immediate vicinity of the $m=1$ OLR, i.e. from $\vert R-R_1\vert\lesssim H_\mathrm{p}$, and from the corotation region $R\approx\Rp$. 
Denoting these contributions as $T_\mathrm{L,id}$ (Lindblad) and $T_\mathrm{c,id}$ (corotation), respectively, we can thus approximate $T_\mathrm{p,id}\approx T_\mathrm{L,id}+T_\mathrm{c,id}$. We corroborate these expectations in Appendix \ref{sec:disc-on-pl}, where we also provide an analytical calculation of $T_\mathrm{L,id}$.

\begin{figure}
	\begin{center}
	\includegraphics[width=0.49\textwidth]{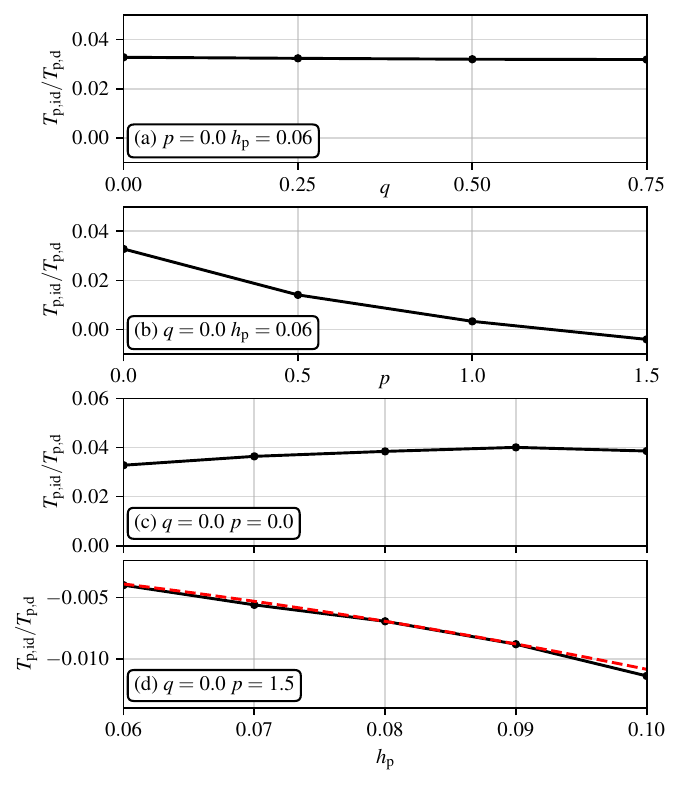}
    \caption{
    Ratio $T_\mathrm{p,id}/T_\mathrm{p,d}$ of the indirect and direct components of the disc torque on the planet, shown as a function of (a) temperature $q$ and (b) surface density $p$ slopes.  In panels (c) and (d) $T_\mathrm{p,id}/T_\mathrm{p,d}$ is shown as a function of $\hp$ for $p=0$ and $p=1.5$, correspondingly. The dashed red curve in panel (d) illustrates the $\hp^{-2}$ scaling. See text for details.
    }
	\label{fig:TiTd}
\end{center}
\end{figure}

\citet{Fairbairn2025a} studied the torque on the planet from the disc and the resultant planetary orbital evolution using semi-analytical linear theory \citep{Miranda2019II,Fairbairn2022}, as described in Section \ref{sec:linear}. We used their approach to numerically compute both $T_\mathrm{p,id}$ and $T_\mathrm{p,d}$ and plot their ratio in Figure \ref{fig:TiTd} as a function of different disc parameters: temperature (a) and $\Sigma$ (b) slopes $q$ and $p$, respectively, and the disc aspect ratio $\hp$, for $q=0$ and two values of $p=0$ (c) and $p=1.5$ (d). 

The most important thing to note in this Figure is that $T_\mathrm{p,id}$ contributes only a small amount to the net torque compared to $T_\mathrm{p,d}$, at the level of several per cent. The ratio $T_\mathrm{p,id}/T_\mathrm{p,d}$ is rather insensitive to the temperature slope $q$ (at least for $p=0$), see panel (a), but shows a strong dependence on the density slope $p$ (for $q=0$), even changing sign for $p\gtrsim 1$, see panel (b). When plotted as a function of $\hp$, this ratio does not vary much for $p=0$; but for $p=1.5$ it is much smaller in amplitude and shows a strong dependence on $\hp$, roughly $\vert T_\mathrm{p,id}/T_\mathrm{p,d}\vert \propto\hp^{2}$.

To shed light on these trends, we note that in a Keplerian disc with $p=1.5$, one expects the corotation torque to be absent, $T_\mathrm{c,id}\to 0$ \citep{GT80}. This is discussed in more detail in Appendix \ref{sec:disc-on-pl}, and is illustrated in Figure \ref{fig:dTidR}b. As a result, for this $p$ the indirect torque on the planet is provided only by the Lindblad contribution, $T_\mathrm{p,id}\approx T_\mathrm{L,id}$, which is known to be independent of $\cs$ or $\hp$ to zeroth order \citep[see][]{GT79,GT80}, as underlined by the discussion in Appendix \ref{sec:disc-on-pl} and our analytical estimate (\ref{eq:TLid-est}). As a result, it follows from equation (\ref{eq:Tpd_est}) that $\vert T_\mathrm{p,id}/T_\mathrm{p,d}\vert\propto\hp^{2}$ for $p=1.5$, in agreement with Figure \ref{fig:TiTd}d.

On the other hand, in a disc with $p=0$ the indirect corotation torque $T_\mathrm{c,id}$ is non-zero and significantly complicates the interpretation of the resultant trends \citep[a similar conclusion has been reached by][]{Crida2025}. We find that in such a disc $T_\mathrm{c,id}$ provides a dominant contribution to $T_\mathrm{p,id}$ (e.g. see Figure \ref{fig:dTidR}a), strongly exceeding that of $T_\mathrm{L,id}$. For that reason, $T_\mathrm{p,id}$ has a higher amplitude for the $p=0$ disc as compared to the $p=1.5$ disc (the changing relative contributions of $T_\mathrm{c,id}$ and $T_\mathrm{L,id}$ cause the particular dependence on $p$ shown in Figure \ref{fig:TiTd}b). However, it has to be remembered that the magnitude of the corotation torque is a strong function of the potential softening, which itself depends on $\hp$ \citep[our linear calculations here adopt a softening length of $0.3\hp$, following][]{Fairbairn2025a}; hence, there is some ambiguity in the actual amplitude of $T_\mathrm{c,id}$ when $p\neq 1.5$ (see Appendix \ref{sec:disc-on-pl}). Also, as a result of this dependence of $T_\mathrm{c,id}$ on the potential softening length, we find that the ratio $\vert T_\mathrm{p,id}/T_\mathrm{p,d}\vert$ ends up being roughly independent\footnote{In other words, we find that $T_\mathrm{p,id}\propto \hp^{-2}$ approximately for $p=0$, which is somewhat coincidentally caused by partial cancellation of $T_\mathrm{c,id}\propto \hp^{-1}$ and $T_\mathrm{L,id}\propto \hp^0$ (they have opposite signs), see Appendix \ref{sec:disc-on-pl}.} of $\hp$ for $p=0$, see Figure \ref{fig:TiTd}c.  

The small relative contribution of $T_\mathrm{p,id}$ to the full planetary torque $T_\mathrm{p}$, obvious from the results presented in Figure \ref{fig:TiTd}, is typical when the disc is not strongly perturbed by the planet and is smooth around its orbit (a situation realized for $\Mp\lesssim \Mth$), in which case the power-law $\Sigma_0$ profile (\ref{eq:PLs}) is appropriate. In this case, most of the torque acting on the planet arises due to the strong direct gravitational coupling at $m\sim\hp^{-1}$ Lindblad resonances located near the planetary orbit \citep{GT80}, and $T_\mathrm{p,id}$ pales in comparison. As a result, the calculations of the rate of Type I planetary migration \citep{Ward1997} should not be strongly affected when the indirect force due to disc is neglected  altogether \citep{Fairbairn2025a}.


\subsection{High mass planets}
\label{sec:id-contrib-p-high}


The situation is different for more massive planets that can carve deep gaps around their orbits. Gap opening strongly suppresses the direct contribution of $m\gtrsim \hp^{-1}$ Lindblad resonances to $T_\mathrm{p}$ as $\Sigma$ is dramatically lowered at their locations. In this case the indirect Lindblad torque contribution $T_\mathrm{L,id}$ at the $m=1$ OLR would grow in relative significance (compared to $T_\mathrm{p,d}$) as its location $R_1=2^{2/3}\Rp$ is far from $\Rp$ and $\Sigma$ there will not be suppressed nearly as much as around $\Rp$. In the most extreme case, when $\Sigma$ is strongly reduced everywhere except the $m=1$ OLR, only $T_\mathrm{L,id}$ and the $m=1$ direct Lindblad contribution $T_\mathrm{L,d}$ would contribute to the torque on the planet $T_\mathrm{p}$. 

As we show in Appendices \ref{sec:L-d} and \ref{sec:L-id}, these two torque contributions are comparable in magnitude but different in sign, with $T_\mathrm{L,id}\approx -0.74\,T_\mathrm{L,d}$ (see equations (\ref{eq:TLd-est}), (\ref{eq:TLid-est}), Appendices \ref{sec:L-d}-\ref{sec:L-verif} and Figure \ref{fig:TLiTLdrat}). As a result, the full torque on the planet coming from the $m=1$ OLR $T_\mathrm{p}=T_{\mathrm{L,d}}+T_{\mathrm{L,id}}$ is roughly 4 times smaller in magnitude than just the direct component $T_{\mathrm{L,d}}$ This implies that in studies of disc-planet coupling in high-$\Mp$ regime, the indirect torque of the disc on the planet can be very significant and should be explicitly taken into account in the calculation of $T_\mathrm{p}$. This is definitely necessary when the planet is capable of opening a gap or cavity around its orbit and its migration switches to the Type II regime \citep{Ward1997}. An analogous conclusion was reached by \citet{RegalyVorobyov2017} and \citet{Crida2025} through direct numerical simulations.


\section{Discussion}
\label{sec:disc}


The main goal of our study was to understand the quantitative impact of excluding the IT from the calculation of disc-planet interaction in a non-inertial, astrocentric frame. Note that we are not questioning the conceptual necessity of including the IT in this frame --- its presence unambiguously follows from the equations of motion in the non-inertial frame (see Section \ref{sec:math}). We are in agreement with \citet{Crida2025} that the IT associated with a particular component of the system (e.g. a planet or a disc) should be included/excluded when the direct gravitational effect of that component is included/excluded in the calculation. But here we are interested in the actual quantitative differences arising when the IT is dropped from the calculation.

When the indirect term is neglected directly in the equations of motion, as in equations (\ref{eq:gasEoMRgen-noIT}) and (\ref{eq:plEoMRgen-noIT}), the  $m=1$ component of the perturbation pattern induced by the planet in the disc gets modified.  The corresponding change of $\delta\Sigma$ can reach considerable magnitude in the outer disc --- around $20\%$ of the amplitude of the full perturbation in the disc at $R\sim 5\Rp$, see Figure \ref{fig:2d_diff_sigma_IT_no_IT_001} --- even in the linear regime; in the inner disc the deviations are negligible. In more nonlinear situations, with higher mass planets, the deviations spread beyond the $m=1$ harmonic, as Figure \ref{fig:fourier_comparison_0.3} illustrates.   

These deviations largely occur far from the planet and their presence does not appreciably affect the amount of angular momentum that a planet excites in the disc by its gravity directly, as illustrated by the $T_\mathrm{d}$ curves in Figures \ref{fig:AM-lin} and \ref{fig:AMF}. But when one also explicitly includes the planetary indirect force in the torque calculation, the differences become dramatic, with the corresponding indirect torque $T_\mathrm{id}$ showing a characteristic oscillatory behavior with growing amplitude. While this behavior may seem peculiar and unphysical, the explicit omission of the indirect torque would actually result in a non-conservation of angular momentum of the planet-driven density waves, as demonstrated in  Section \ref{sec:AMF}: the AMF profile of the density wave would still exhibit such oscillations as long as the IT was accounted for in the calculation of the disc perturbation. Thus, to ensure conservation of the wave angular momentum expressed through equation (\ref{eq:Tsplit1}) one {\it must} account for the indirect torque $T_\mathrm{id}$. 

Somewhat ironically, the same indirect torque oscillations also reflect the {\it non-conservation} of the angular momentum of the full star-planet-disc system in the non-inertial astrocentric frame, which is evident from equation (\ref{eq:Tpdsum}), keeping in mind that the stellar angular momentum is zero in this frame. This non-conservation arises as a result of the torque exerted by the indirect force $M\aid$ ($M$ is the total mass of the system including $M_\star$) applied to the barycenter of the system, as shown in \citet{Rafikov2025}.

We would like to re-iterate that the indirect torque oscillations should not be confused with the torque wiggles studied in \citet{Cimerman2024b} or the torque oscillations in circumbinary discs explored in \citet{Cimerman2024a}. These phenomena arise because of the direct gravitational force of the planet coupling with the global planet-driven spiral density wave on scales $\gtrsim \Rp$. On the other hand, the indirect torque oscillations are due to the indirect force coupling to the same density wave on global scales. The difference is easily seen in the behavior of the oscillation amplitudes --- the indirect torque oscillations grow with $R$, whereas torque wiggles decay with $R$ \citep{Cimerman2024b}. The two types of torque oscillations actually co-exist, as can be noticed in Figures \ref{fig:dTdR}-\ref{fig:AMF} (the first torque wiggle around $2.2\Rp$ can be seen in both $\de T/\de R$ and $F_J$ curves). 

Despite the divergent oscillations at large $R$ exhibited by both the wave AMF and the full integrated torque $T=T_\mathrm{d}+T_\mathrm{id}$, the deposition torque density $\de T_\mathrm{dep}/\de R$ --- the difference between the two --- is well behaved and decays to zero as $R\to \infty$ (inasmuch as numerical artefacts allow one to determine it), see Figure \ref{fig:bary_astro_deposition}. Moreover, the profile of $\de T_\mathrm{dep}/\de R$ is largely independent of whether it is computed in the non-inertial frame, barycentric frame, or not accounting for the IT at all, at least for low (sub-$\Mth$) mass planets. This is good news for studies of gap opening by planets, since the behavior of $\de T_\mathrm{dep}/\de R$ --- the amount of angular momentum absorbed by the disc material per unit radius --- is the key driver of disc evolution resulting from planet-disc coupling. Thus, even studies that disregard the IT altogether should still provide a reliable description of gap opening by the low mass planets. 

However, the situation should change in the high-$\Mp$ regime, when a planet opens a deep gap, if the direct excitation torque density $\de T_\mathrm{d}/\de R$ ends up being very different between the calculations including and disregarding the IT due to the planet. In particular, \citet{RegalyVorobyov2017} found that inclusion of the IT  in their simulations decreases the width of the gap carved by a planet. This may be a consequence of the lower amplitude of $\delta\Sigma_1$ (see Figures \ref{fig:fourier_comparison}c \& \ref{fig:fourier_comparison_0.3}c) at $m=1$ OLR when the IT is included, weakening disc-planet coupling and reducing both the total direct planetary torque exerted on the outer disc and the deposited torque.

Similarly, the amount of angular momentum absorbed (or lost) by a low-mass (sub-$\Mth$) planet changes weakly, at the level of several per cent, if the IT due to the disc is turned off, as we demonstrated in Section  \ref{sec:id-contrib-p}. Such small changes to the torque on the planet $T_\mathrm{p}$ are expected to only slightly alter the planetary migration rate in the Type I regime. In this study we considered only circular planets, but previously \citet{Fairbairn2025a} showed that even for a non-circular planet, its eccentricity evolution does not change dramatically when the IT is switched off since $T_\mathrm{p}$ changes very little. However, for high mass planets, capable of opening deep gaps around their orbit, the situation may be very different, as the contribution to the torque coming from the normally dominant high $m\sim \hp^{-1}$ Lindblad resonances close to the planet \citep{GT80} gets suppressed. As a result, the indirect component of the disc torque on the planet excited at $m=1$ OLR starts playing a very important role (see Section \ref{sec:id-contrib-p-high}). Thus, the indirect component of this torque may significantly affect the rate of Type II migration for massive planets. 

Summarizing this discussion, we find that neglecting the IT in disc-planet interaction calculations often results in only small differences that can be tolerated. In particular, for low mass planets, in the linear regime, neither the disc evolution (i.e. gap opening) nor the orbital evolution of the planet (its radial migration) should change by much. Nevertheless, fully accounting for the IT due to some massive component of the system is still highly recommended whenever the direct gravitational effect of that component is included, even if just for self-consistency \citep{Crida2025}. And in some cases, e.g. for higher mass planets, the IT should play an important quantitative role in evolution of the system (e.g. see Section \ref{sec:id-contrib-p-high}). 

Some results stemming from calculations that fully and self-consistently employ the IT may look surprising at first --- the oscillatory behavior of the indirect torque and AMF in runs including the IT, non-conservation of global angular momentum of the system, etc. --- and should be interpreted with care. Whenever in doubt regarding the origin of a particular outcome of a calculation in the astrocentric frame, we recommend verifying that outcome by repeating the calculation in the inertial, barycentric frame.


\subsection{Boundary conditions for simulations}
\label{sec:BCs}


Use of a non-inertial, moving reference frame in studying the disc-planet interaction can have other consequences, besides from introducing the indirect force. One such implication has to do with the boundary conditions (BCs) used in simulations of disc-planet coupling. In cylindrical polar coordinates, traditionally used for modeling protoplanetary discs, BCs are imposed at a fixed distance from the grid center. In the case of an astrocentric frame this is done at a fixed $R$, whereas modeling in the barycentric frame imposes BCs at a fixed barycentric distance $r$. 

In reality, far from the planet ($R\gg\Rp$) the disc tends to orbit around the system barycenter, so the natural axisymmetric BCs should be applied at fixed $r$. Applying them at fixed $R$, as routinely done in astrocentric simulations, leads to azimuthally-varying deviations of the fluid velocities at the outer boundary relative to the true BCs, the amplitude of which scales $\propto\Mp/\M_\star$. As a result, a parasitic perturbation is forced in the flow at the outer boundary which may propagate through the whole domain and contaminate the results of a simulation. This problem has been previously noted\footnote{Earlier, \citet{KP93} noted that `The indirect term contributes not only in the body of the disk, but also through a term that is strongly dependent in a complicated way on the exact treatment of the disk edge.'} in \citet{Teys2017} and a recipe for dealing with it was outlined in \citet[][see their Appendix B]{Dempsey2021}. We independently rediscovered this issue and recommend mitigating it in simulations by applying the appropriate azimuthally-varying outer BCs at fixed $R$.   

The situation is opposite in the inner disc, where for $R\ll\Rp$ the disc tends to orbit the central star and not the system barycenter. In this case the standard inner BCs at fixed $R$ would naturally work well for simulations in the astrocentric frame. However, calculations using the barycentric frame would need to adjust the inner BCs accordingly, otherwise the inner boundary would force an unwanted perturbation in the computational domain, which gets larger for a more massive planet and for a smaller radius $r$ of the inner boundary.


\subsection{Comparison with existing literature}
\label{sec:lit}


There is an extensive body of literature on disc-planet coupling, both numerical and analytical. Some of these studies include the IT, while others do not; we do not name specific examples as the lists would be too long both ways. In some cases the inclusion or neglect of the IT is clearly specified, but often it is not mentioned explicitly. We support \citet{Crida2025} in urging everyone to explicitly state whether the IT due to each component of the system under study is included (or not) in the calculations performed in the astrocentric frame.

When the IT is explicitly neglected, this approach is usually justified on the grounds of simplicity, or because it does not make large differences in the outcomes, although typically no quantitative assessment of these differences is provided. \citet{KP93} and \citet{Miranda2019I} also motivated the neglect of the IT for practical purposes, since its inclusion raises compatibility issues with the radiative boundary conditions adopted in their linear calculations. As our results demonstrate, in many cases (predominantly for low-mass planets) the neglect of the IT leads to only minor changes in the evolution of the disc and planetary orbit, corroborating the results of earlier studies not accounting for the IT. 

We are not aware of any studies that tried to explicitly compare the perturbation patterns induced by a planet with or without the planetary IT, as we do in Section \ref{sec:pert}. The planetary torque on the disc has been previously computed by a number of authors in different ways. For example, \citet{Bate2003}, \citet{Duffell2012}, \citet{Cimerman2024b} do not include the planetary IT in their calculations of the $\Sigma$ perturbation, while \citet{Kley2012}, \citet{Arza18} do. However, in all of these cases, only the direct torque component was computed, so that $T=T_\mathrm{d}$ was assumed. To the best of our knowledge, prior to our work no other study of planet-disc coupling accounted for the indirect torque. It was considered only in the context of modeling discs in stellar binaries (cataclysmic variables) by \citet{Ju2016,Ju2017}. However, since such discs lie {\it interior} to the orbit of the perturber, the indirect torque plays a negligible role (in the absence of permanent eccentricity) and exhibits no prominent oscillations (see Section \ref{sec:analytical}).

Similarly, not many planet-disc studies computed the AMF $F_J$ in a global setting, and when this was done, the IT was neglected altogether (e.g. \citealt{Cimerman2021}, \citealt{Cimerman2024b}). The work of \citet{Zia2023} represents an interesting exception: they include the planetary IT in the calculation of the disc perturbation in the astrocentric frame and also compute $F_J$. They comment on `some oscillations in the radial profiles of the AMF' in the outer disc (see their Fig. 5), which, in light of our results, should be interpreted as the AMF oscillations reflecting the effect of the indirect torque $T_\mathrm{id}$. \citet{Brown2025} also accounted for the planetary IT in the calculation of the $\Sigma$ perturbation and computed $F_J$ but only in the disc interior to the planetary orbit, where no interesting effects due to the indirect forces are expected.

Studies of planetary orbital evolution typically neglected indirect forces \citep{KP93,Pap2000}. The situation changed recently when \citet{Fairbairn2025a} computed planetary migration and eccentricity damping timescales, both with and without the disc IT. Working in the linear regime ($\Mp\ll\Mth$), they found these timescales to change at most by $\sim 10\%$ when the disc IT was not included. Later, \citet{Crida2025} supported this conclusion with direct numerical simulations. They also found that the disc IT plays a more substantial role in planet migration for high $\Mp$, in the Type II migration\footnote{\citet{RegalyVorobyov2017} also found different Type II migration rates when allowing or prohibiting the star to move in their simulations. However, they ascribe this difference to a different gap structure between the two cases, which is a consequence of the planetary IT rather than the disc IT.} regime. All of these findings are in full agreement with our (semi-)analytical results in Section \ref{sec:mig} and Appendix \ref{sec:disc-on-pl}.


\section{Summary}
\label{sec:sum}


In this work we explored the impact of indirect forces, arising due to non-inertial motion of the central star, on different aspects of disc-planet coupling. Focusing on the case of a planet on a circular orbit in a non-self-gravitating disc, we considered modifications induced by fully accounting for the IT to (1) the perturbation pattern driven by the planetary gravity in the disc, (2) the calculation of the torque exerted by a planet on the disc, and (3) the torque exerted by the disc on a planet. Our key findings are briefly summarized below.
\begin{itemize}

\item In the linear regime ($\Mp\ll\Mth$) only the $m=1$ Fourier harmonic of the perturbation pattern $\delta\Sigma$ changes when one does not account for the IT. The difference in the calculation with the IT included increases with $R$, reaching a mean magnitude of $\sim 20\%$ of the maximum amplitude of the full perturbation pattern at $5\Rp$. Meanwhile, in the inner disc this difference is negligible.

\item At higher planetary masses, $\Mp\gtrsim 0.3\Mth$, the deviation caused by the IT propagates to higher-$m$ harmonics of $\delta\Sigma(R,\phi)$.

\item When the planetary IT is included in the calculation of $\delta\Sigma(R,\phi)$, the indirect force must be explicitly accounted for in calculation of the torque exerted on the disc. Without this the angular momentum of the planet-driven density wave is not conserved.

\item The indirect torque $T_\mathrm{id}$ on the disc due to the planetary IT has an oscillatory structure, with radial periodicity determined by the global structure of the density wake. The amplitude of $T_\mathrm{id}$ oscillations typically increases with $R$. The angular momentum of the full star-planet-disc system is not conserved, as expected in a non-inertial frame.

\item At the same time, the deposition torque density, which determines disc evolution and gap opening, varies little whether or not the planetary IT is included.

\item For low-mass planets, $\Mp\lesssim\Mth$, the torque on the planet $T_\mathrm{p}$ exerted by the disc perturbation $\delta\Sigma$ is only weakly affected (at the level of several per cent) when the IT due to the disc is included in the calculation. As a result, Type I migration rates should be only weakly affected by non-inclusion of disc IT. For high-mass, gap-opening planets, the disc IT  plays a more important role, potentially changing the rate of Type II migration compared to the case without the disc IT. 

\end{itemize}

Overall, in many cases the inclusion of IT leads to only minor changes in the disc and planetary orbital evolution. Nevertheless, we strongly recommend  properly including it in the calculations of disc-planet coupling. We also encourage all researchers to explicitly state whether they do or do not include the IT in their studies of disc-planet interaction.


\section*{Acknowledgements}


We are grateful to Lev Arzamasskiy, Alexandros Ziampras, Scott Tremaine and Jim Stone for useful discussions, and to an anonymous referee for helpful suggestions. R.R.R. acknowledges financial support from the STFC grant ST/T00049X/1 and the IBM Einstein Fellowship at the IAS. A.J.D. was supported by NASA through the NASA Hubble Fellowship grant
\#HST-HF2-51553.001 awarded by the Space Telescope Science Institute, which is operated by the Association of Universities for Research in Astronomy, Inc., for NASA, under contract NAS5-26555. C.W.F acknowledges support provided by the Friends of the IAS Members fund.

\section*{Data Availability}
The data underlying this article will be shared on reasonable request to the corresponding author.




\bibliographystyle{mnras}
\bibliography{references} 



\appendix


\section{Oscillations of Fourier components of  $\delta\Sigma$ in $\Mp=0.3\Mth$ case}
\label{sec:fourier_oscillations}


Here we explain the non-monotonic behavior of $\Sigma_m$ featuring multiple nulls in Figure \ref{fig:fourier_comparison_0.3}, illustrating a mildly non-linear case with $\Mp=0.3\Mth$. We show that this behavior has its origin in the azimuthal broadening of the planet-driven density wave caused by nonlinear effects \citep{GR01}, which is evident in Figure \ref{fig:fourier_comparison_0.3}a.

\begin{figure}
	\begin{center}
	\includegraphics[width=0.49\textwidth]{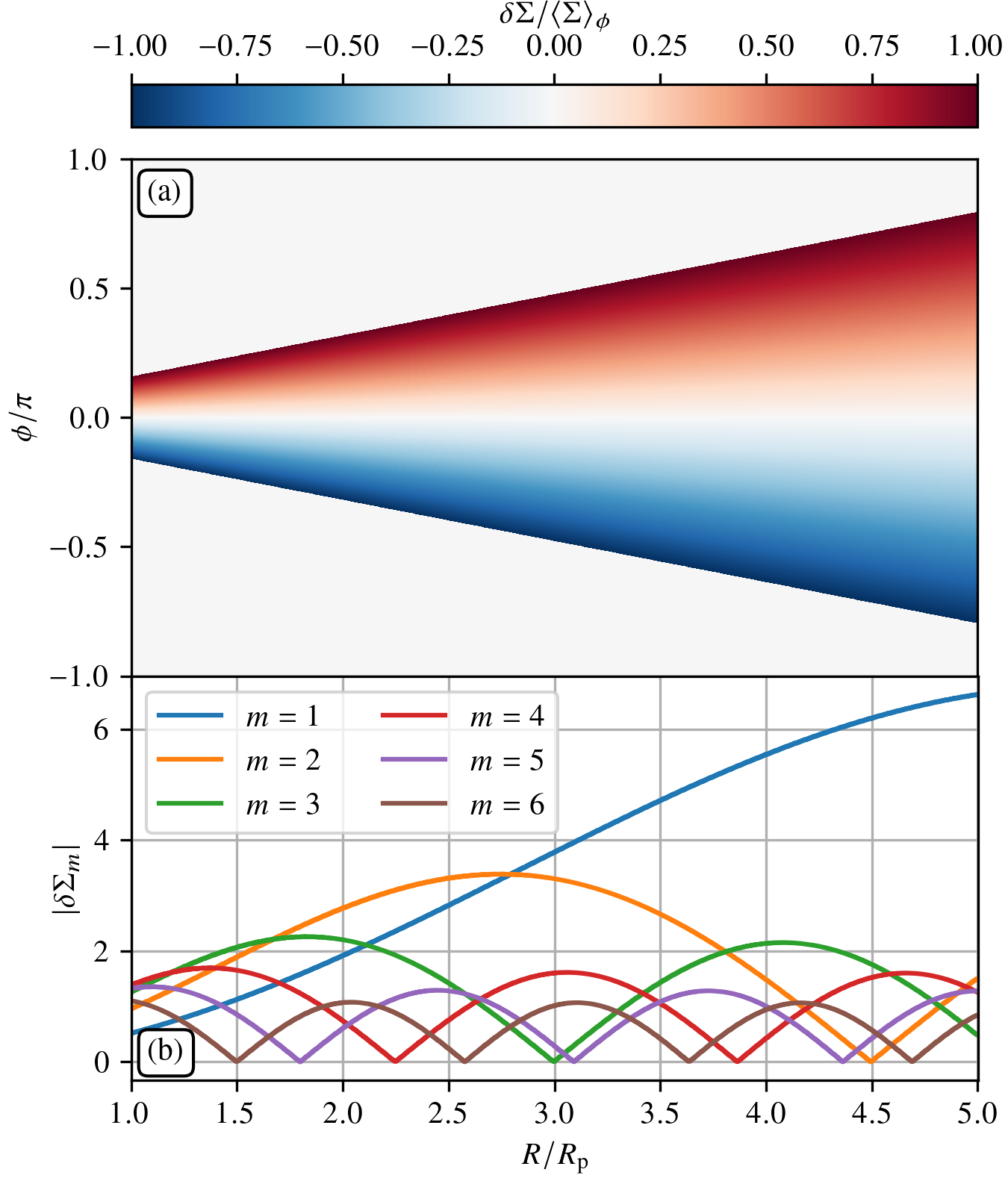}
	\caption{
(a) Cartesian $R-\phi$ map of the surface density perturbation which spreads azimuthally (profile width $w(R)$ increases) as the distance from the center increases, illustrating the toy model (\ref{eq:Nanzatz}).  (b) Low-$m$ Fourier harmonics of the perturbed $\Sigma$ exhibiting periodic nulls.
}
	\label{fig:toy_sigma}
	\end{center}
\end{figure}

Let us consider a $\Sigma$ perturbation in a self-similar form 
\begin{align}
\delta\Sigma(R,\phi)=A(R)f\left(\frac{\phi-\phi_0(R)}{w(R)}\right),
\label{eq:anzatz}
\end{align}
where $A(R)$ is the wave amplitude, $f(z)$ is a function determining the overall azimuthal wave profile, $w(R)$ is a (radially-varying) width of the profile, and the function $\phi_0(R)$ determines the overall winding of the wave in the $(R,\phi)$ plane. We will assume that $f(z)$ is non-zero only within a range of $z$, which we set to be $(-z_0,z_0)$, with $z_0$ being some constant, without loss of generality. We will also assume that the azimuthal wave profile does not cover the full $2\pi$ range in $\phi$, which requires the profile width to obey $z_0w(R)<\pi$.

Using these assumptions and the ansatz (\ref{eq:anzatz}), the Fourier mode equation (\ref{eq:Sig_m}) can be re-written as
\ba
\delta\Sigma_m=\frac{Aw}{2\pi}\e^{-\I\, m\phi_0}\int_{-z_0}^{z_0}f(z)\,\e^{-\I\, mwz}\,\de z,
\label{eq:Sig_m1}
\ea  
so that 
\ba
\vert\delta\Sigma_m\vert=\frac{Aw}{2\pi}\left\vert\int_{-z_0}^{z_0}f(z)\,\e^{-\I\, mwz}\,\de z\right\vert.
\label{eq:Sig_m2}
\ea  

The nonlinear evolution of the density wave (which is essentially an acoustic wave) eventually results in the formation of an N-wave profile, which can be modeled as 
\begin{align}
f(z)=z,~~~|z|<z_0,
\label{eq:Nanzatz}
\end{align}
and zero otherwise. The evolution of the planet-driven density wave profile in Figure \ref{fig:fourier_comparison_0.3}a can be very roughly described by this model, with a radially increasing $w(R)$. In this case, evaluating the integral in 
(\ref{eq:Sig_m2}), one finds
\begin{align}  
|\delta\Sigma_m(R)|=\frac{Aw}{\pi}\frac{\left[(mwz_0)^2+1\right]^{1/2}}{(mw)^2}\left|\sin\left(mwz_0-\psi_0\right)\right|,
\label{eq:Nwave}
\end{align}  
where $\tan\psi_0=mwz_0$. As the profile width $w(R)$ increases with $R$, $\delta\Sigma_m$ must exhibit radial oscillations with higher (radial) frequency for higher $m$. For $mwz_0\gg 1$, $\psi_0\to \pi/2$ and the nulls of $\delta\Sigma_m$ would lie at radii where $mw(R)z_0=\pi(k+1/2)$, with $k$ being an integer. Note that the shape of the wave in the $(R,\phi)$ plane is irrelevant, since $|\delta\Sigma_m(R)|$ does not depend on $\phi_0(R)$.

We illustrate this behavior in Figure \ref{fig:toy_sigma} using a model of the wave in accordance with equation (\ref{eq:Nanzatz}) where we adopt $\phi_0(R)=0$, constant $A(R)$ and linearly increasing $w(R)$, see panel (a). In panel (b) we show $\delta\Sigma_m(R)$ for $m=1,\dots,6$, illustrating their oscillatory behaviour with $R$. As expected, $\delta\Sigma_m$ feature more nulls for higher $m$, which is also noticeable in Figure \ref{fig:fourier_comparison_0.3}b.


\section{Torque density expressions}
\label{sec:torque-deriv}


Here we provide details of the derivation for the indirect torque density due to the planetary potential acting on the disc (see equation (\ref{eq:Tid})). Using the polar coordinate system $(R,\phi)$ with the angle $\phi$ counted from the instantaneous $\bfRp$ direction, we can re-write equation (\ref{eq:dTdRid}) as
\begin{align}
\frac{\de T_\mathrm{id}(R)}{\de R} = -\pi G\Mp\frac{R^2}{\Rp^2}\Sigma_1^s(R),
\label{eq:dTdRid1}
\end{align}
where $\Sigma_1^s$ is the $m=1$ sin-coefficient $\Sigma_m^s(R)$ of the Fourier expansion for $\Sigma(R,\phi)$:
\ba
\Sigma_m^s(R)=\pi^{-1}\int_0^{2\pi}\Sigma(R,\phi)\,\sin (m\phi)\,\de\phi.
\label{eq:sin_coef}
\ea  
Upon identifying that $\Sigma_m^s=-2\,\IM\,\Sigma_m$, where $\Sigma_m$ is the coefficient (defined by equation (\ref{eq:Sig_m})) of the Fourier expansion $\Sigma(R,\phi)=\sum_{m=-\infty}^\infty \Sigma_m(R)\exp(im\phi)$, expression  (\ref{eq:dTdRid1}) necessarily reduces to equation (\ref{eq:Tid}).

For completeness, we also provide the expression for the torque density contribution due to the direct component of the planetary gravity, see equation (\ref{eq:dTdRd}): 
\begin{align}
\frac{\de T_\mathrm{d}(R)}{\de R} & = G \Mp R \int_0^{2\pi}
\frac{\Sigma(R,\phi)\, R \Rp\sin\phi \, \de\phi}
{\left(R^2+\Rp^2-2R \Rp\cos\phi\right)^{3/2}}
\nonumber\\
 & = \frac{G \Mp R^2\Rp}{[\max(R,\Rp)]^3} \int_0^{2\pi} \frac{\Sigma(R,\phi)\, \sin\phi\, \de\phi}{\left(1+\alpha^2-2\alpha\cos\phi\right)^{3/2}}.
\label{eq:dTdR1}
\end{align}
where we introduced $\alpha=\min(R,\Rp)/\max(R,\Rp)<1$. Using the relation 
\begin{align}
\frac{\sin\phi}{\left(1+\alpha^2-2\alpha\cos\phi\right)^{3/2}} = 
\alpha^{-1}\sum\limits_{m=1}^\infty mb_{1/2}^{(m)}(\alpha)\sin (m\phi),
\label{eq:identity}
\end{align}
\citep[see equation (A4) of][]{Cimerman2024a} with 
\begin{align}
    b^{(m)}_{1/2} (\alpha) = \frac{1}{\pi}
                    \int\limits_{0}^{2\pi}
                        \frac{\cos (m \theta)\,\de \theta}{(1 + \alpha^2 - 2 \alpha \cos \theta)^{1/2}}
    \label{eq:lap_def}
\end{align}
being the Laplace coefficients \citep[e.g.][]{MurrayDermott}, equation (\ref{eq:dTdR1}) becomes
\begin{align}
\frac{\de T_\mathrm{d}(R)}{\de R}  & =\pi G \Mp\alpha^\zeta \sum\limits_{m=1}^\infty m \,b_{1/2}^{(m)}(\alpha) \,\Sigma_m^s(R)
\nonumber\\
& =-2\pi G \Mp\alpha^\zeta \sum\limits_{m=1}^\infty m \,b_{1/2}^{(m)}(\alpha) \,\IM\,\Sigma_m(R).
\label{eq:Td-gen}
\end{align}
Here we introduced an indicator exponent $\zeta=\Theta(\Rp-R)$ with $\Theta(z)$ being the Heavyside step-function, making the expression (\ref{eq:Td-gen}) applicable in both the inner ($R<\Rp$, $\zeta=1$) and the outer ($R>\Rp$, $\zeta=0$) parts of the disc. In a somewhat different form this expression has been previously derived in \citet{Cimerman2024b}; its generalization in an inertial, barycentric frame has been obtained in \citet{Cimerman2024a} in the context of circumbinary discs.

Equations (\ref{eq:Tid}) and (\ref{eq:Td-gen}) are derived without making any approximations and are fully general.  They allow one to compute the torque acting on the disc if the behavior of the Fourier coefficients $\Sigma_m(R)=\delta\Sigma_m(R)$ is known, e.g. from simulations or based on some theoretical considerations. In general, $\Sigma_m$ can be functions of not only $R$ but also time $t$.


\section{Indirect disc torque on a planet}
\label{sec:disc-on-pl}


Here we provide details on the calculation of the indirect component of the disc torque on the planet, as described in Section \ref{sec:id-contrib-p}.


\subsection{Lindblad vs corotation torque separation}
\label{sec:torque-sep}

\begin{figure}
	\begin{center}
	\includegraphics[width=0.49\textwidth]{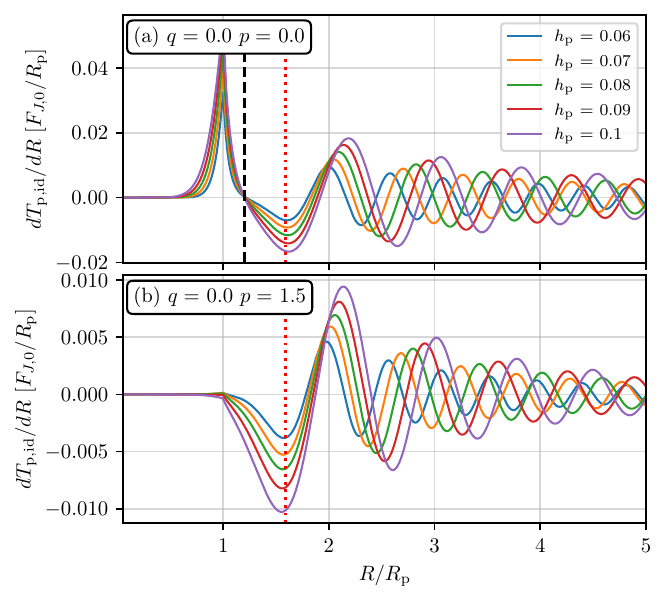}
    \caption{Radial density 
$\de T_\mathrm{p,id}/\de R$ of the indirect disc torque acting on the planet, plotted for $q=0$, (a) $p=0$ and (b) $p=1.5$ and several values of $\hp$. Dashed vertical line in panel (a) marks the boundary of the regions where the torque is assigned either to corotation ($R<1.2\Rp$) or Lindblad resonance ($R>1.2\Rp$) for $p\neq 1.5$. Note that the peak at corotation is absent for $p=1.5$. For $p=0$ the height of the peak at corotation depends on $\hp$. Dotted vertical line marks the position of the outer Lindblad resonance $R_1$. Note that excitation of the Lindblad torque is a rather non-local process.}
	\label{fig:dTidR}
\end{center}
\end{figure}

First, we look at the separation of the torque on the planet induced by the indirect potential of the disc $T_\mathrm{p,id}$ into the corotation and Lindblad contributions, $T_\mathrm{c,id}$ and $T_\mathrm{L,id}$ respectively.  In Figure \ref{fig:dTidR} we plot the differential contribution $\de T_\mathrm{p,id}/\de R$ to the indirect disc torque onto the planet from different radii, computed based on equation (\ref{eq:Tpid}), for two values of $p$ and several values of $\hp$. 

For $p=0$ one can see both the oscillatory part of $\de T_\mathrm{p,id}/\de R$ roughly outside of the $m=1$ OLR and the peak at the corotation ($R=\Rp$). Note that the corotation peak is not particularly narrow, and to obtain $T_\mathrm{c,id}$ we integrate $\de T_\mathrm{p,id}/\de R$ up to the somewhat arbitrarily selected radius $R=1.2\Rp$, roughly where  $\de T_\mathrm{p,id}/\de R$ changes sign. The remaining integral of $\de T_\mathrm{p,id}/\de R$, from $R=1.2\Rp$ and above, is then identified with $T_\mathrm{L,id}$. This procedure is different from the one employed in \citet{Fairbairn2025a} to split the direct planetary torque on the disc into the corotation and Lindblad contributions --- measuring the jump of the disc AMF $F_J$ at the corotation resonance. This is because here we are interested in the indirect effect of the disc on the planet, and not vice versa.

\begin{figure}
	\begin{center}
	\includegraphics[width=\columnwidth]{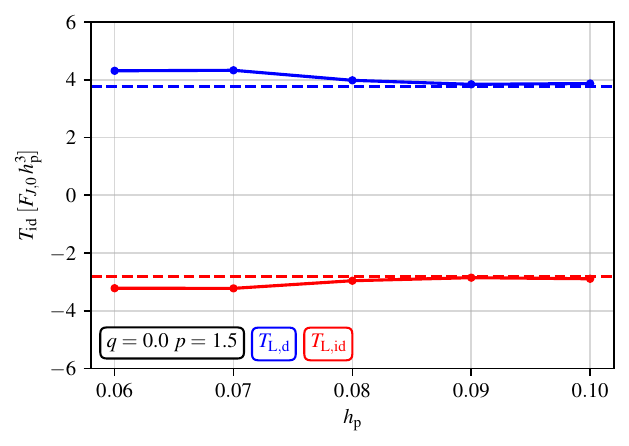}
    \caption{
    Direct (blue) and indirect (red) $m=1$ Lindblad contributions $T_\mathrm{L,d}$ and $T_\mathrm{L,id}$
    of the disc torque on the planet. Dashed lines show the analytical predictions (\ref{eq:TLd-est}) and (\ref{eq:TLid-est}), while the dots (connected by solid lines) show the results of our numerical linear calculation for $p=1.5$, $q=0$ as a function of $\hp$. 
    }
	\label{fig:TLiTLd}
\end{center}
\end{figure}

\begin{figure}
	\begin{center}
	\includegraphics[width=\columnwidth]{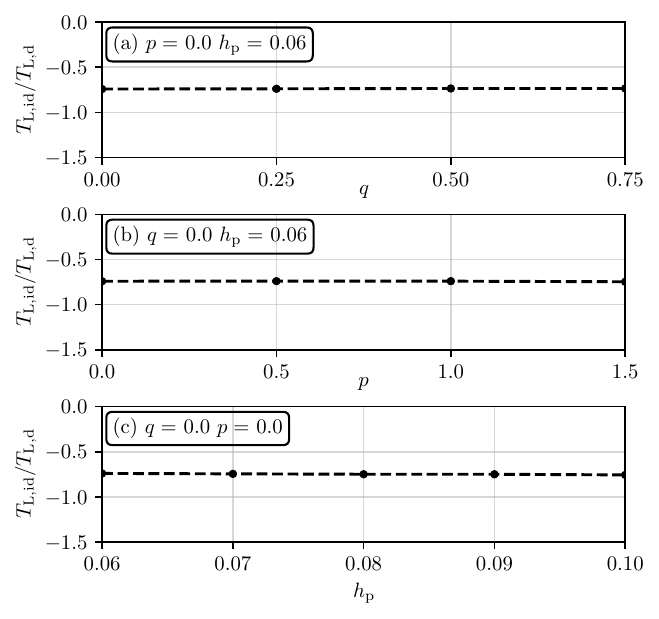}
    \caption{
        Ratio of the indirect $T_\mathrm{L,id}$ and direct  $T_\mathrm{L,d}$ components of the $m=1$ Lindblad contributions as a function of (a) $q$, (b) $p$, and (c) $\hp$. Note that $T_\mathrm{L,id}/T_\mathrm{L,d}$ stays close to the analytically predicted value of $\approx -0.74$.
    }
	\label{fig:TLiTLdrat}
\end{center}
\end{figure}

Note that for $p=0$ the peak value of $\de T_\mathrm{p,id}/\de R$ at corotation increases as $\hp$ decreases, which is caused by the decreasing softening length (linearly scaling with $\hp$) of the planetary potential that contributes to the density perturbation at the corotation. For that reason the strength of the indirect corotation contribution $T_\mathrm{c,id}$ should depend on $\hp$, with the scaling $T_\mathrm{c,id}\propto \hp^{-1}$ \citep{GT79}, which is indeed the case (see Section \ref{sec:id-contrib-p}).

For $p=1.5$ (Figure \ref{fig:dTidR}b) the peak at the corotation is absent, in agreement with the theoretical expectation \citep{GT79,GT80} for the lack of corotation torque in a $p=1.5$ disc, which has no vortensity gradient. Thus, in this case we assign the full indirect torque due to the disc to $T_\mathrm{L,id}$. 

Figure \ref{fig:dTidR} also clearly shows that in the framework of our linear calculation the inner disc provides essentially no contribution to $T_\mathrm{p,id}$, consistent with analytical expectations. Indeed, $T_\mathrm{p,id}$ is determined by the $m=1$ Fourier harmonic $\delta\Sigma_1=\Sigma_1$ of the surface density perturbation (see equation (\ref{eq:Tpid1})), which is extremely small in the inner disc for $\Mp\ll\Mth$ according to Figure \ref{fig:fourier_comparison}b. We note in this regard that \citet{Crida2025} did find a considerable non-zero (albeit rapidly oscillating) contribution to $T_\mathrm{p,id}$ from the inner disc. This may be somehow related to the higher $\Mp=0.08\Mth$ that they considered; our $\Mp=0.3\Mth$ calculation clearly shows small but non-zero $\Sigma_1$ in the inner disc (see Figure \ref{fig:fourier_comparison_0.3}b), likely caused by the growth of disc eccentricity in the inner disc, see Section \ref{sec:pert-nonlin}.

Both panels of Figure \ref{fig:dTidR} make it clear that excitation of the Lindblad torque $T_\mathrm{L,id}$ is not concentrated in the immediate vicinity of $R_\mathrm{L}$, as expected from the classical analysis of \citep{GT79,GT80} and mentioned in Section \ref{sec:id-contrib-p}. In reality, near and outside the $m=1$ OLR, $\de T_\mathrm{p,id}/\de R$ exhibits slowly decaying oscillations and $T_\mathrm{L,id}$ converges to its final value rather far\footnote{The non-locality of tidal coupling at Lindblad resonances is also the reason behind the negative torque density phenomenon studied in \citet{Dong2011} and \citet{RP12}.} from the planet. The origin of these oscillations is analogous to that of the binary torque oscillations in circumbinary discs \citep{Cimerman2024a} and can again be traced to the global structure of the $m=1$ perturbation in the disc (see Figure \ref{fig:2d_diff_sigma_IT_no_IT_001}). 


\subsection{Direct $m=1$ Lindblad torque $T_\mathrm{L,d}$}
\label{sec:L-d}


Next, we demonstrate how to obtain analytical estimates of $T_\mathrm{L,id}$ and $T_\mathrm{L,d}$ by separating the Lindblad torque into the indirect and direct components. We do this for a disc with $p=1.5$, for which there is no ambiguity in separating the torque into the Lindblad and corotation components (see Section \ref{sec:torque-sep}) and a direct comparison with the semi-analytical linear calculation can be provided (see Section \ref{sec:L-verif}).

To compute the direct $m=1$ Lindblad torque component $T_\mathrm{L,d}$ we will use the fact that the direct planetary torque on the disc is equal to the direct disc torque on the planet (e.g. see equation (\ref{eq:Tpd})). This allows us to use the torque calculation machinery developed by \citet{GT79}, see equations (A7)-(A10) in the Appendix of that work. However, we need to generalize their results, since \citet[][using their notation]{GT79} computed the torque exerted by the potential perturbation component $\varphi_1$ on the surface density perturbation $\sigma_1$ generated by the same potential perturbation $\varphi_1$; note that here the subscripts in $\sigma_1$ and $\varphi_1$ refer to first-order perturbations, not $m=1$ Fourier components. In our case, $\sigma_1$ is due to the full perturbing potential --- i.e. the sum of the direct and indirect planetary potentials -- but the calculation of $T_\mathrm{L,d}$ must account only for the torque due to the direct planetary potential acting on this $\sigma_1$. 

Let us consider a general case when the $\Sigma$ perturbation is caused by a perturbing potential (its $m$-th azimuthal harmonic) $\Phi_\Sigma$, while the force enacting the torque is the gradient of some other potential $\Phi_\mathrm{T}$. This means that in the equation (A7) of \citet{GT79} we associate $\varphi_1$ with $\Phi_\mathrm{T}$, while $\sigma_1$ is induced by $\Phi_\Sigma$. Propagating this through the derivation in the equations (A7)-(A10) of \citet{GT79}, we find the corresponding Lindblad torque at $m$-th OLR to be 
\begin{align}
T_\mathrm{L,m}=-m\pi^2\left\{\Sigma\left(R\frac{\de D}{\de R}\right)^{-1}\Psi\left[\Phi_\Sigma\right]\Psi\left[\Phi_\mathrm{T}\right]\right\}_{R_\mathrm{L}} \,,
\label{eq:TpL}
\end{align}
where $D=\Omega^2-m^2(\Omega-\Omegap)^2$ and we have defined 
\begin{align}
\Psi\left[\Phi\right]=R\frac{\de\Phi}{\de R}+\frac{2\Omega}{\Omega-\Omegap}\Phi.
\label{eq:Psi}
\end{align}
In what follows we confine ourselves to $m=1$.

In our case the perturbation of $\Sigma$ is caused by the full planetary potential, including its indirect part, $\Phi_\Sigma = -(G\Mp/\Rp)[b^{(1)}_{1/2}(\alpha)-\alpha]$, $\alpha=R/\Rp$ for $m=1$ \citep{GT80}. To find the direct part $T_\mathrm{L,d}$ of the $m=1$ Lindblad torque, we take $\Phi_\mathrm{T}$ to be given by only the direct part of this expression, $\Phi_\mathrm{T}=-(G\Mp/\Rp)b^{(1)}_{1/2}(\alpha)$. Plugging this into (\ref{eq:TpL}) and setting $m=1$, $p=1.5$ (to express $\Sigma(R_1)$ through $\Sigma_\mathrm{p}$ and enable comparison in Figure \ref{fig:TLiTLd}) and $R_\mathrm{L}=R_1=2^{2/3}\Rp$, we obtain \begin{align}
T_\mathrm{L,d}\approx 3.78 F_{J,0}\hp^3\,,
\label{eq:TLd-est}
\end{align}
which is independent of $\hp$, in lieu of the definition (\ref{eq:FJ0}). This is in agreement with the classical theory \citep{GT79,GT80} predicting that the torque excited at an individual Lindblad resonance should be independent of $\cs$ and $\hp$.


\subsection{Indirect Lindblad torque $T_\mathrm{L,id}$}
\label{sec:L-id}


To find $T_\mathrm{L,id}$ we cannot use the symmetry property of the direct torques that we exploited for computing $T_\mathrm{L,d}$, see the discussion around equations (\ref{eq:Tpid}) and (\ref{eq:Tpid2}). Instead, we note that our equation (\ref{eq:Tpid1}) looks identical to the initial torque expression (A7) of \citet{GT79} if we set $\varphi_1=\Phi_\mathrm{T}=(G\Mp\Rp/R^2)$; this implies that the subsequent result (\ref{eq:TpL}) can still be used to compute $T_\mathrm{L,id}$ if we adopt this `effective' form of $\Phi_\mathrm{T}$ \citep{Fairbairn2025a}, keeping $\Phi_\Sigma$ as above. Performing this procedure for a $p=1.5$ disc, we arrive at 
\begin{align}
T_\mathrm{L,id}\approx -2.8 F_{J,0}\hp^3\,,
\label{eq:TLid-est}
\end{align}
again independent of $\hp$.


\subsection{Lindblad torque estimates: verification}
\label{sec:L-verif}


We verify our analytical calculation of $T_\mathrm{L,d}$ and $T_\mathrm{L,id}$ in Figure \ref{fig:TLiTLd} using our numerical linear calculations for $p=1.5$ (when $T_\mathrm{c,id}=0$) and $q=0$, finding good agreement with the analytical results given by equations (\ref{eq:TLd-est}) and (\ref{eq:TLid-est}). The variation of these torque components with $\hp$ is rather minor, as expected. The slight differences for lower $\hp$ are likely caused by the non-locality of disc-planet coupling at the $m=1$ OLR, as noted earlier. Our present results, as well as those of \citet{KP93}, \citet{RP12} and \citet{Fairbairn2025a}, suggest that the simplifying assumption of the locality of Lindblad torque excitation near the OLR, which was adopted in \citet{GT79,GT80}, may result in some minor inaccuracies.

In Figure \ref{fig:TLiTLdrat} we show the numerically-computed ratio $T_\mathrm{L,id}/T_\mathrm{L,d}$ of the indirect to direct torque components at the $m=1$ OLR as a function of $q$, $p$ and $\hp$. Knowledge of this ratio is relevant for understanding planet migration in the Type II regime, see Section \ref{sec:id-contrib-p}.  Analytical calculations in Sections \ref{sec:L-d} and \ref{sec:L-id} yield 
$T_\mathrm{L,id}/T_\mathrm{L,d}\approx -0.74$ (see equations (\ref{eq:TLd-est}) and (\ref{eq:TLid-est})), and this is what we indeed see in Figure \ref{fig:TLiTLdrat}, regardless of $q$, $p$ or $\hp$.


\bsp	
\label{lastpage}
\end{document}